\pdfoutput=1
\documentclass[fleqn]{SCIS}
\def\co{$^{12}$CO}
\def\13co{$^{13}$CO}
\def\c18o{C$^{18}$O}
\def\ci{[C\,{\sc i}]}
\def\cii{[C\,{\sc ii}]}
\def\r{$\rho$}
\def\nh{n(H$_2$)}

\def\r{$\rho$}
\def\arcsec{\hbox{$^{\prime\prime}$}}
\def\arcmin{\hbox{$^{\prime}$}}

\def\cm2{cm$^{-2}$}
\def\cc{cm$^{-3}$}
\def\kms{km s$^{-1}$}

\def\nh3{NH$_3$}
\def\h2{H$_2$}
\def\nh{n(H$_2$)}

\def\r{$\rho$}
\usepackage[normalem]{ulem}
\usepackage[utf8]{inputenc}
\usepackage[T1]{fontenc}

\newcommand{\aap}{A\&A}
\newcommand{\apj}{ApJ}
\newcommand{\apjl}{ApJL}
\newcommand{\apjs}{ApJS}
\newcommand{\mnras}{MNRAS}
\newcommand{\nat}{Nature}

\newcommand{\aj}{AJ}

\newcommand{\pasj}{Publ. Astron. Soc. Jpn.}

\def\araa{ARA\&A}  

\begin{document}

\ensubject{subject}

%%%%%%%%%%%%%%%%%%%%%%%%%%%%%%%%%%%%%%%%%%%%%%%%%%%%%%%
%%% Authors do not modify the information below
%%% ????????????????
%%% ??????????, ????????????{}, ???????????????????
%Letter to the Editor??Article%??????
\ArticleType{Article}%??Article
\SpecialTopic{SPECIAL TOPIC: }%???????
\Year{2025}
\Month{TBC}
\Vol{TBC}
\No{1}
\DOI{??}
\ArtNo{000000}
\ReceiveDate{TBC}
\AcceptDate{TBC}
%\OnlineDate{January 1, 2016}
%%%%%%%%%%%%%%%%%%%%%%%%%%%%%%%%%%%%%%%%%%%%%%%%%%%%%%%

%%% title: ????
%%%   \title{title}{title for citation}
\title{Large scale mapping of \ci\ and the \ci-to-CO transition in \r\ Ophiuchus molecular cloud}

%%% Corresponding author: ???????

\author[1,2]{Jifeng Xia}{}
\author[3]{Ningyu Tang}{{nytang@ahnu.edu.cn}}
\author[4]{Thomas G. Bisbas}{}
\author[5]{Chen Wang}{}
\author[6]{Gan Luo}{}
\author[1]{Sihan Jiao}{}
\author[1,2]{Xin Lv}{}
\author[4]{\\Xuejian Jiang}{} 
\author[11]{Donghui Quan}{}
\author[1]{Jinzeng Li}{}
\author[7]{Paul F. Goldsmith}{}
\author[8,9]{Gary A. Fuller}{}
\author[10,1]{Di Li}{{dili@mail.tsinghua.edu.cn}}
\footnote{*Corresponding author:nytang@ahnu.edu.cn, dili@mail.tsinghua.edu.cn}

%%% Author information for page head.

\AuthorMark{Jifeng Xia}
\AuthorCitation{Jifeng Xia, Ningyu Tang, Thomas G. Bisbas, et al}

%%%   \address[number]{Address, City {\rm Postcode}, Country}
\address[1]{National Astronomical Observatories, Chinese Academy of Sciences, Beijing 100101, China}
\address[2]{University of Chinese Academy of Sciences, Beijing 100049, China}
\address[3]{Department of Physics, Anhui Normal University, Wuhu, Anhui 241002, China}
\address[4]{Research Center for Astronomical Computing, Zhejiang Laboratory, Hangzhou 311100, China}
\address[5]{Institute of Astronomy and Information, Dali University, Dali, 671003, China}
\address[6]{Institut de Radioastronomie Millimetrique, 300 rue de la Piscine, 38400, Saint-Martin d’Hères, France}
\address[7]{Jet Propulsion Laboratory, California Institute of Technology, 4800 Oak Grove Drive, Pasadena, CA 91109, USA}
\address[8]{Jodrell Bank Centre for Astrophysics, Oxford Road, The University of Manchester, Manchester M13 9PL, UK}
% \address[9]{I. Physikalisches Institut, Universit¨at zu K¨oln, Z¨ulpicher Straße 77, 50937 Cologne, Germany}

\address[9] {I. Physikalisches Institut, Universität zu Köln, Zülpicher Straße 77, 50937 Cologne, Germany}
\address[10] { New Cornerstone Science Laboratory, Department of Astronomy, Tsinghua University, Beijing 100084, China}
\address[11] { Department of Physics, Xi’an Jiaotong-Liverpool University, Suzhou 215123, China}
%%% Abstract. ??
\abstract{Atomic carbon (\ci) is a key species in the carbon chemistry of the interstellar medium (ISM).  Using the Submillimeter Wave Astronomy Satellite (SWAS), we conducted a \ci ($^3$P$_1$--$^3$P$_0$) 492 GHz survey covering approximately 4 deg$^2$ of the L1688 and L1689 regions in the \r\ Oph molecular cloud, achieving a spatial resolution of 4.25$\arcmin$.  The derived \ci\ column densities, N(\ci), range from 4.85 $\times$ 10$^{14}$ cm$^{-2}$ to 6.29 $\times$ 10$^{17}$ cm$^{-2}$, corresponding to an abundance ratio N(\ci)/N(\h2) of 2.24$\times$ 10$^{-7}$ to 2.39$\times$ 10$^{-4}$, with a median value of 1.8$\times$ 10$^{-5}$. Combining observations with photodissociation region (PDR) modeling, we find that \ci\ abundance varies less than CO in regions with UV intensity G$_0$ $> 16$ and N(H$_2$) $<$ 4.6 $\times$ 10$^{21}$ cm$^{-2}$, suggesting \ci\ is a more reliable tracer of molecular hydrogen in low-density, high-radiation environments where the \ci-to-CO transition occurs. Utilizing \ci\ as direct \h2\ tracer, the CO-dark gas fraction is estimated to be 0.43 , meaning that 43\% of the total cloud mass will be missed by conventional calculation based on CO observations but can be calibrated by \ci\ emission. The \ci\ line widths are systematically broader  than those of $^{13}$CO, possibly due to contributions from atomic carbon. These findings provide key insights into Galactic \ci\ emission and the carbon cycle evolution in the interstellar medium. Future high-sensitivity \ci ($^3$P$_1$--$^3$P$_0$) surveys with the Chinese Survey Space Telescope (CSST) will significantly advance our understanding of the carbon cycle evolution.}

%%% Keywords. ?????


\maketitle

\begin{multicols}{2}
\section{Introduction}\label{sec:intro}

%% General introduction to Carbon  
As one of the most abundant elements in the universe, ranking fourth after hydrogen, helium, and oxygen, carbon serves as a cornerstone of interstellar chemistry. Observations of atomic carbon (\ci) and carbon monoxide (CO) offer crucial diagnostics for probing the physical and chemical conditions of the interstellar medium (ISM). The \ci\  ($^3P_1 \rightarrow ^3P_0$) fine-structure transition (which has a rest frequency of 492.161 GHz) and the CO ($J=1\rightarrow 0$) rotational line (rest frequency of 115.271 GHz) exhibit critical densities of $1.0\times 10^3$ \cc\ ( assuming temperature of 50 K) and $3\times 10^2$ \cc, respectively. 
Both transitions can be easily excited through colliding with H$_2$ in low-to-intermediate density gas and serve as efficient coolants in the ISM. Due to their high abundance, \ci\, and CO are usually used as experimental tracers of cold molecular gas \cite{1978ApJ...222..881G, 1991ApJ...377..192H}.
%(Goldsmith et al.\ 1978, Hollenbach et al.\ 1991). 

The \ci\ is produced through the recombination of \cii\ with electrons and the photodissociation of CO molecules by ultraviolet (UV) photons (E $>$ 11.09 eV). It is primarily depleted via photoionization by higher-energy UV photons (E $>$ 11.3 eV), as well as through chemical reactions that lead to the formation of CO. In the plane-parallel photodissociation region (PDR) model, the balance between production and depletion confines \ci\ to a narrow layer between the outer C\,{\sc ii} and inner CO zones, where UV flux maintains a low carbon ionization rate \cite{1985ApJ...291..722T, 1986ApJS...62..109V, 1988ApJ...334..771V, 2022ARA&A..60..247W}. The strong self-shielding effect enables \h2\ to persist in this narrow layer where CO cannot survive, forming the so-called ``CO-dark" molecular gas region \cite{1992IAUS..150..143V, 2005Sci...307.1292G, 2011A&A...536A..19P, 2013ARA&A..51..207B, 2014A&A...561A.122L, 2015ApJ...798....6F, 2016A&A...593A..42T, 2018ApJS..235....1L, 2018A&A...611A..51R, 2023A&A...675A.145L, 2020ApJ...889L...4L }. Thus, \ci\ emission is one of the preferred candidate to trace the ``CO-dark" molecular gas.

Despite that the plane-parallel PDR model succeeds in describing the Orion Bar PDR  \cite{1985ApJ...291..747T, 1989ARA&A..27...41G, 1993Sci...262...86T, 1994ApJ...422..136T, 1995A&A...294..792H, 1999RvMP...71..173H, 2016Natur.537..207G}, recent observations reveal that \ci\ follows a spatial distribution comparable to CO in nearby giant molecular clouds (e.g., Orion, DR15, W51, Ophiuchi, RW38) — and maintains significant abundance even in regions with high visual extinction \cite{2002ApJS..139..467I, 2013ApJ...774L..20S, 2001ApJ...558..176O, 1999sf99.proc...88A, 2003ApJ...589..378K, 2021ApJ...914L...9Y, 2021PASJ...73..174I}. 
%(Orion: Ikeda et al. 2002; Shimajiri et al. 2013, DR15: Oka et al. 2001, W51: Arikawa et al. 1999, Ophiuchi: Kamegai et al. 2003, Yamagishi 2021,  RCW38: Izumi et al. 2021). 
These results are inconsistent with predictions from the plane-parallel PDR model.

The widespread distribution of \ci\ emission can be explained through two  physical mechanisms. The first is enhanced cosmic ray flux. In regions with elevated cosmic-ray ionization rates, the efficient dissociation of CO molecules by cosmic rays leads to the formation of extended \ci\ zones into denser regions in plane-parallel PDR models \cite{2015ApJ...803...37B,2017ApJ...839...90B,2015MNRAS.450.4424B}. The second explanation is a clumpy PDR. The clumpy structure of molecular clouds  provides an alternative framework, where the fractal nature of the interstellar medium - characterized by a hierarchy of dense clumps following a power-law size distribution - naturally accounts for both the ubiquity of \ci\ and the observed chemical heterogeneity \cite{1988ApJ...332..379S, 1989ARA&A..27...41G, 1990ApJ...365..620B, 1993ApJ...405..216M, 1997A&A...323..953S, 2004A&A...424..887K, 2008A&A...477..547K, 2008A&A...482..197P, 2008A&A...488..623C}.

Recent advances in three-dimensional numerical simulations  further reveal that \ci\  and CO emission lines likely originate from distinct gas components, a consequence of both the three-dimensional density structure of the interstellar medium and the differential response of these molecular tracers to local environmental parameters such as radiation field and turbulent conditions \cite{2021MNRAS.502.2701B}. Large-scale surveys of both \ci\ and CO toward nearby Giant molecular clouds would provide a chance to further  investigate the origin of \ci.

The \r\ Ophiuchus molecular cloud complex is a nearby region of low-mass and intermediate-mass star formation \cite{2008hsf2.book..351W}. The two primary regions within the complex, L1688 and L1689, are situated at distances of 138.4 ± 2.6 pc and 144.2 ± 1.3 pc, respectively \cite{2018ApJ...869L..33O}. Compared to L1688, L1689 possesses a similar mass but exhibits more subdued star formation activity \cite{2010A&A...518L.102A, 2015A&A...584A..91K}. Additionally, L1688 hosts a B2V star, HD 147889, which is responsible for driving the high UV flux within the L1688 area. These characteristics render the $\rho$ Ophiuchus molecular cloud complex an ideal site for studying the evolution of carbon chemistry under varying star formation conditions \cite{2020A&A...638A..74L}.

Earlier observations of the L1688 cloud and its surrounding extended regions around $\rho$ Ophiuchi A have been conducted through integrated intensity mapping of the \ci\ emissions \cite{2003ApJ...589..378K, 2021ApJ...914L...9Y}. These studies calculated column densities at several representative positions. They provided further insight into the chemical evolution of the region, suggesting that both PDR and time-dependent chemistry play important roles in the distribution of neutral carbon and CO.

In this paper, we present the large-scale \ci\ 492 GHz survey toward   \r\ Ophiuchus molecular cloud including L1688 and L1689  region with  Submillimeter Wave Astronomical Satellite (SWAS) \cite{2000ApJ...539L..77M}. 

Section \ref{sec:data} describes datasets of \ci, CO, and dust observations.  Analysis and results about intensity of UV radiation, column density and spatial distribution of different species are presented in section \ref{sec:results}.  Finally, the discussion and summary are shown in section \ref{sec:discussion} and \ref{sec:summary}, separately. 

\section{Data}\label{sec:data}
\subsection{\ci\ observations}

The lower fine-structure transition of atomic carbon, $\ci$ ($^3$P$_1 \rightarrow {}^3$P$_0$) with rest frequency of 492.161 GHz can be observed from the ground. However, making maps of extended regions from the ground can be particularly difficult due to the need for a large spatial switch in an unstable atmosphere. As a NASA explorer mission, the SWAS  was capable of observing the \ci\ line with about 1 MHz frequency resolution (corresponding to a velocity resolution of 0.76 km s$^{-1}$ at 492 GHz), beam size of   3.5\arcmin $\times$ 5.0\arcmin, and system temperature of $\sim 170$ K \cite{2000ApJ...539L..77M}.  

Between February 1999 and March 2002, SWAS completed a 4 square degree \ci\ map toward the Ophiuchi region with 4345 pointings spaced 1.6\arcmin\ apart (PI: Di Li). The data were convolved to a Gaussian beam of 4.25\arcmin\ Full Width Half Maximum, close to the angular resolution of SWAS. The observations are oversampled for the telescope beam, and reach rms noise level of 0.24 K per 0.76 km s$^{-1}$ channel. The Pixels in the data cube are spaced by 2.125\arcmin\ on the sky. A main beam efficiency of 0.9 is adopted to convert the data from antenna temperature ($T\rm_A$) to brightness temperature ($T\rm_{MB}$) \cite{2000ApJ...539L..77M}.

\subsection{CO data}
The $^{12}$CO (J=1-0; 115.271 GHz) and the $^{13}$CO  (J=1-0; 110.207 GHz) archival data were obtained using the SEQUOIA 32-element focal plane array at the 14-m Five College Radio Astronomy Observatory (FCRAO) telescope in New Salem, Massachusetts, employing on-the-fly (OTF) mapping techniques. 
These observations are part of COMPLETE (Coordinated Molecular Probe Line Extinction and Thermal Emission) Survey of Star-Forming Regions \cite{2006AJ....131.2921R}.  
The original data have a spatial resolution of 44\arcsec\ for $^{12}$CO(1-0) and 46\arcsec\  for $^{13}$CO(1-0),  and a spectral resolution of 0.06 km s$^{-1}$. After correcting for main beam efficiency of 0.5 at 110 Ghz and 0.45 at 115 GHz, the reduced mean rms of brightness temperature is 0.98 K per 0.064 km s$^{-1}$ for $^{12}$CO and 0.33 K per 0.064 km s$^{-1}$ for $^{13}$CO \cite{2006AJ....131.2921R}. 

Between May 2001 and January 2002, we conducted FCRAO observations of the C$^{18}$O (J=1-0; 109.782 GHz) using frequency-switched mode (PI: Di Li). The observations achieved a sensitivity of 0.5 K (in $T\rm_A$) per 0.1 km s$^{-1}$ with a main beam efficiency of ~0.5 at 109 GHz. 

The map center ($\alpha_{2000}$=16:26:23, $\delta_{2000}$=-24:23:02) covers most of the L1688 and L1689 regions. The spatial distributions of CO and its isotopologues are shown in Fig. \ref{fig:skycover}. To keep consistency with the \ci\ data resolution, we convolved the $^{12}$CO, $^{13}$CO and C$^{18}$O (1-0) data to a uniform resolution of 4.25\arcmin. This processing yielded rms noise levels of 0.42 K per 0.06 km s$^{-1}$ channel for $^{12}$CO,  0.13 K per 0.06 km s$^{-1}$ channel for $^{13}$CO and 0.2 per 0.1 \kms\ channel for C$^{18}$O.

\subsection{Archival Dust Temperature and H$_2$ Column Density}

\begin{figure*}[ht!]
\centering
 \includegraphics[width=0.98\textwidth]{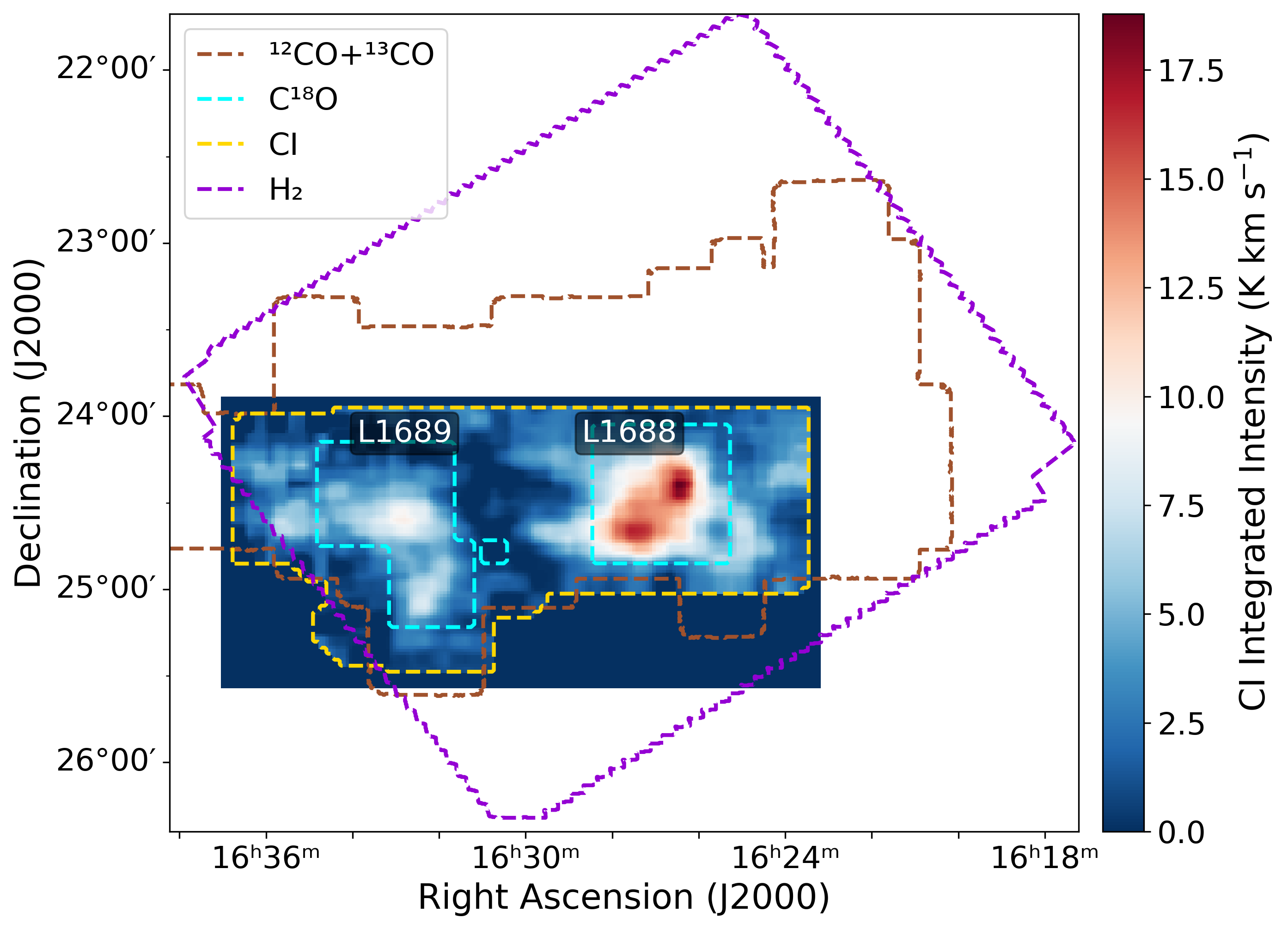}
 \caption{Sky coverage of multi-wavelength observations toward the $\rho$ Ophiuchi cloud. The background is the integrated intensity of \ci\. The line in brown is the contour of the data coverage of $^{12}$CO and $^{13}$CO by FCRAO. The line in light blue is the contour of the data coverage of C$^{18}$O by FCRAO. The line in yellow is the contour of the data coverage of \ci\ by SWAS. The line in purple is the contour of the data coverage of dust emission observed by \textit{Herschel} and used to derive the H$_2$ column density.} 
 \label{fig:skycover}
 \end{figure*}

The dust temperature and \h2\ column density data were obtained from the \textit{Herschel Gould Belt Survey} (HGBS) \cite{2010A&A...518L.102A}, conducted using the \textit{Herschel Space Observatory}. \textit{Herschel}, an ESA mission with NASA participation, carried the Photodetector Array Camera and Spectrometer (PACS) and Spectral and Photometric Imaging Receiver (SPIRE) instruments, which observed in four bands (70, 160, 250, 350, and 500 $\mu$m). For the $\rho$ Oph region, the \h2\ column density and dust temperature maps were derived through spectral energy distribution (SED) fitting, with an original spatial resolution of 37.3\arcsec \cite{2014A&A...562A.138R, 2020A&A...638A..74L}. To facilitate comparison with other datasets, we convolved these maps to a common resolution of 4.25\arcmin. The sky coverage of these data products is shown in Fig. \ref{fig:skycover}.

\section{Analysis and Results}\label{sec:results}

\subsection{Intensity Distribution}
\label{subsec:intg-codark}

\begin{figure*}[ht!]
\centering
 \includegraphics[width=0.95\textwidth]{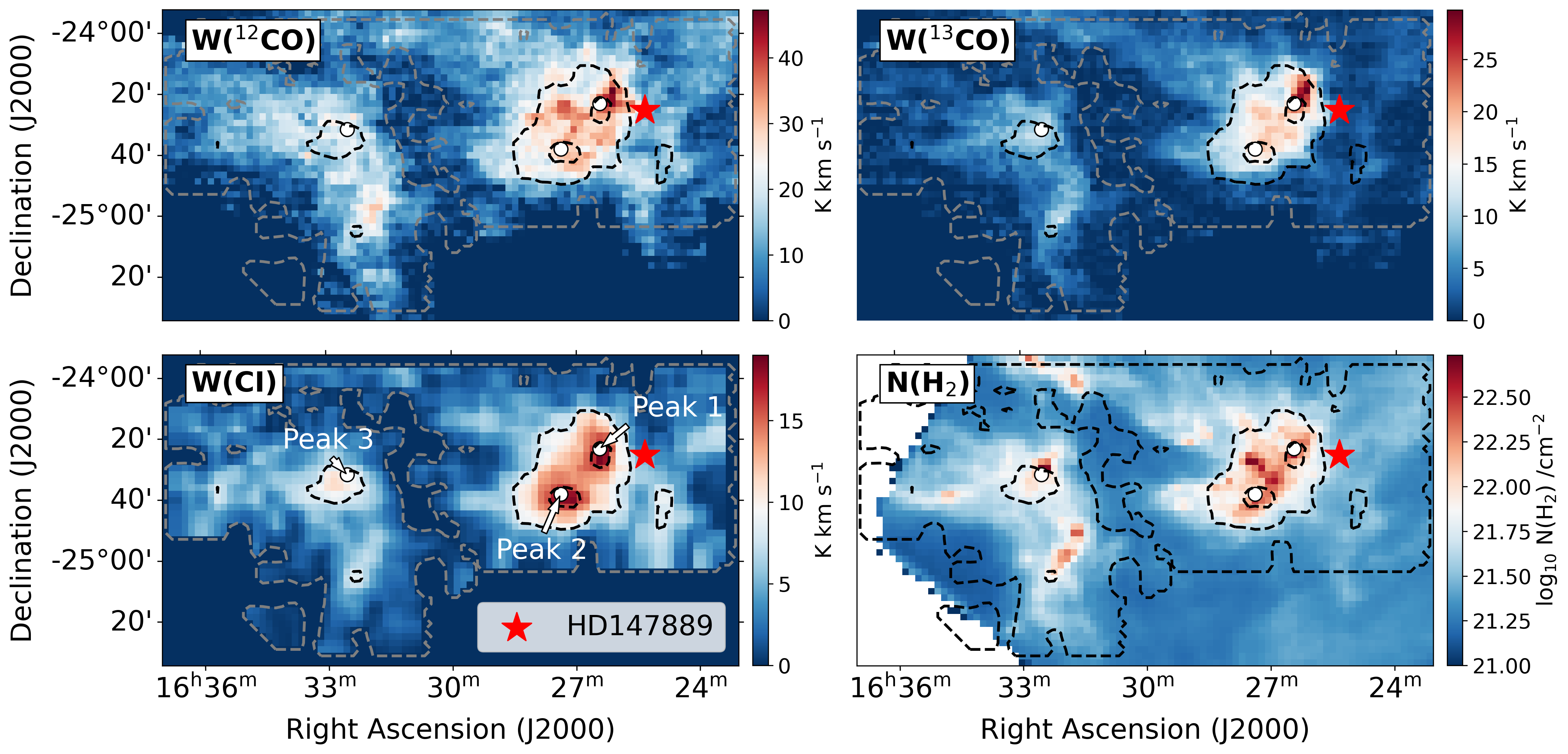}
 \caption{The contours of the total integrated intensity maps of \ci\,($^3$P$_1$ -$^3$P$_0$) emission overlaid on images of four different tracers: $^{12}$CO (integrated from -15\,km\,s$^{-1}$ to 20\,km\,s$^{-1}$), $^{13}$CO  (integrated from  -15\,km\,s$^{-1}$ to 20\,km\,s$^{-1}$), \ci\  (integrated from -15\,km\,s$^{-1}$ to 20\,km\,s$^{-1}$), and H$_2$ column density map. The white contour levels are 0, 10, and 20 K\,km\,s$^{-1}$. The B2V star HD 147889 is indicated by a red star symbol in all panels.}
 \label{fig:cicontour}
 \end{figure*}

\begin{figure*}[ht!]
\centering
 \includegraphics[width=0.98\textwidth]{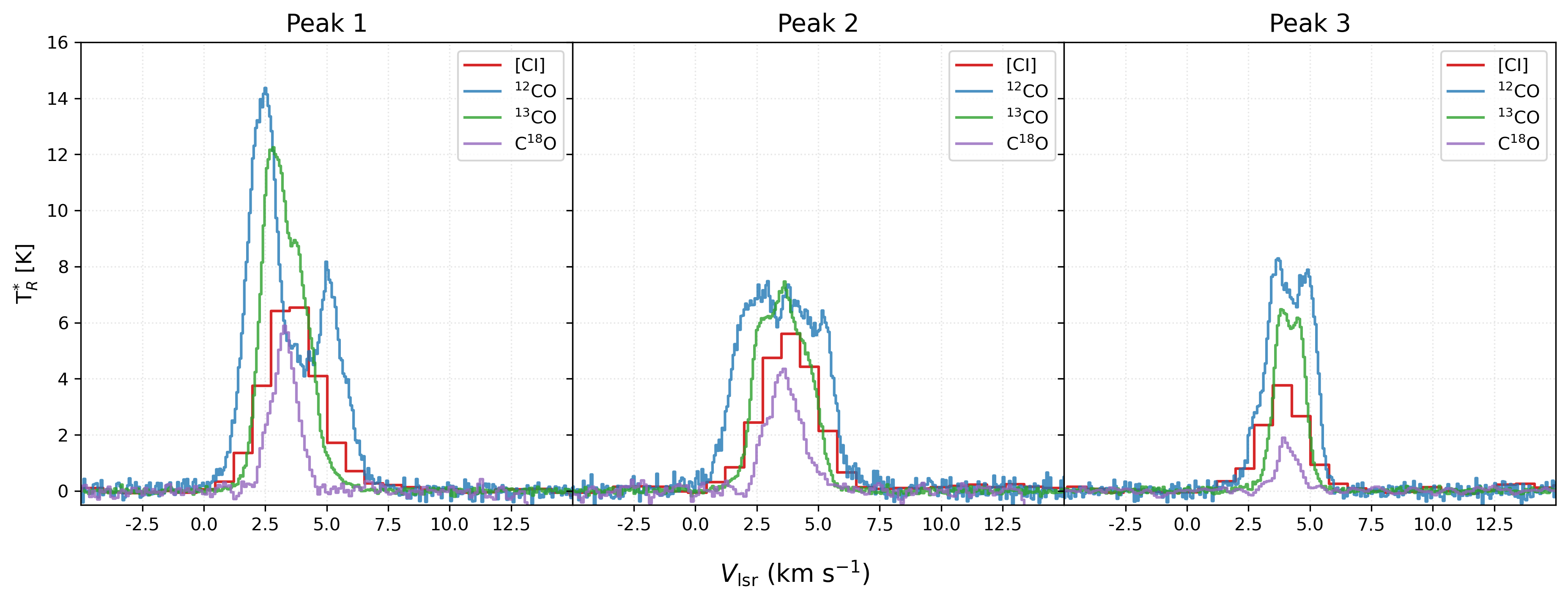}
 \caption{Spectra of \ci\, $^{12}$CO, $^{13}$CO, and C$^{18}$O
 emission with initial velocity resolution toward three selected positions: Peak 1, Peak 2 and Peak 3.}
\label{fig:spectra}
\end{figure*}

Fig. \ref{fig:cicontour} presents the spatial distributions of the integrated intensity of $^{12}$CO, $^{13}$CO and \ci\, along with the H$_2$ column density. In the L1688 region, the \ci\ integrated intensity exhibits two distinct peaks, designated as Peak 1 and Peak 2. Peak 1 ($\alpha_{2000}$=16:26:36, $\delta_{2000}$=-24:25:04) is located in proximity to the massive star HD 147889, while Peak 2 ($\alpha_{2000}$=16:27:33, $\delta_{2000}$=-24:40:05) is offset by approximately 19\arcmin\ from the star. Notably, Peak 1 spatially coincides with the $\rho$ Oph A clump observed in both $^{12}$CO and $^{13}$CO intensity maps, whereas Peak 2 shows no such correspondence. The \h2\ column density, N(\h2), reaches its maximum value near Peak 2 but does not display significant enhancement around Peak 1. The pronounced emissions of $^{12}$CO, $^{13}$CO, and \ci\ at Peak 1 may be attributed to excitation enhancement caused by the nearby massive star HD 147889. This interpretation is consistent with findings from previous studies \cite{2003ApJ...589..378K}. In the L1689 region, the peak position of \ci\ intensity, Peak 3  ($\alpha_{2000}$=16:32:49, $\delta_{2000}$=-24:37:27) is also shown in Fig. \ref{fig:cicontour}. This Peak appears to be in spatial displacement from those of $^{12}$CO and $^{13}$CO, but demonstrates better agreement with the peak location in the N(\h2) map.  The spectra of \ci\ , $^{12}$CO, $^{13}$CO, and C$^{18}$O toward three dense positions (Peak 1, Peak 2 and Peak 3) are presented in Fig. \ref{fig:spectra}. The central velocity of \ci\ emission is consistent with that of the C$^{18}$O emission.

 As shown in  Fig.~\ref{fig:cicontour}, the large-scale distribution of \ci\ extends into more diffuse regions as traced by $^{12}$CO  than those traced by $^{12}$CO. Notably, we detect CO-dark gas that exhibits prominent \ci\ emission but weak CO emission.

\begin{figure*}[ht!]
\centering
 \includegraphics[width=0.95\textwidth]{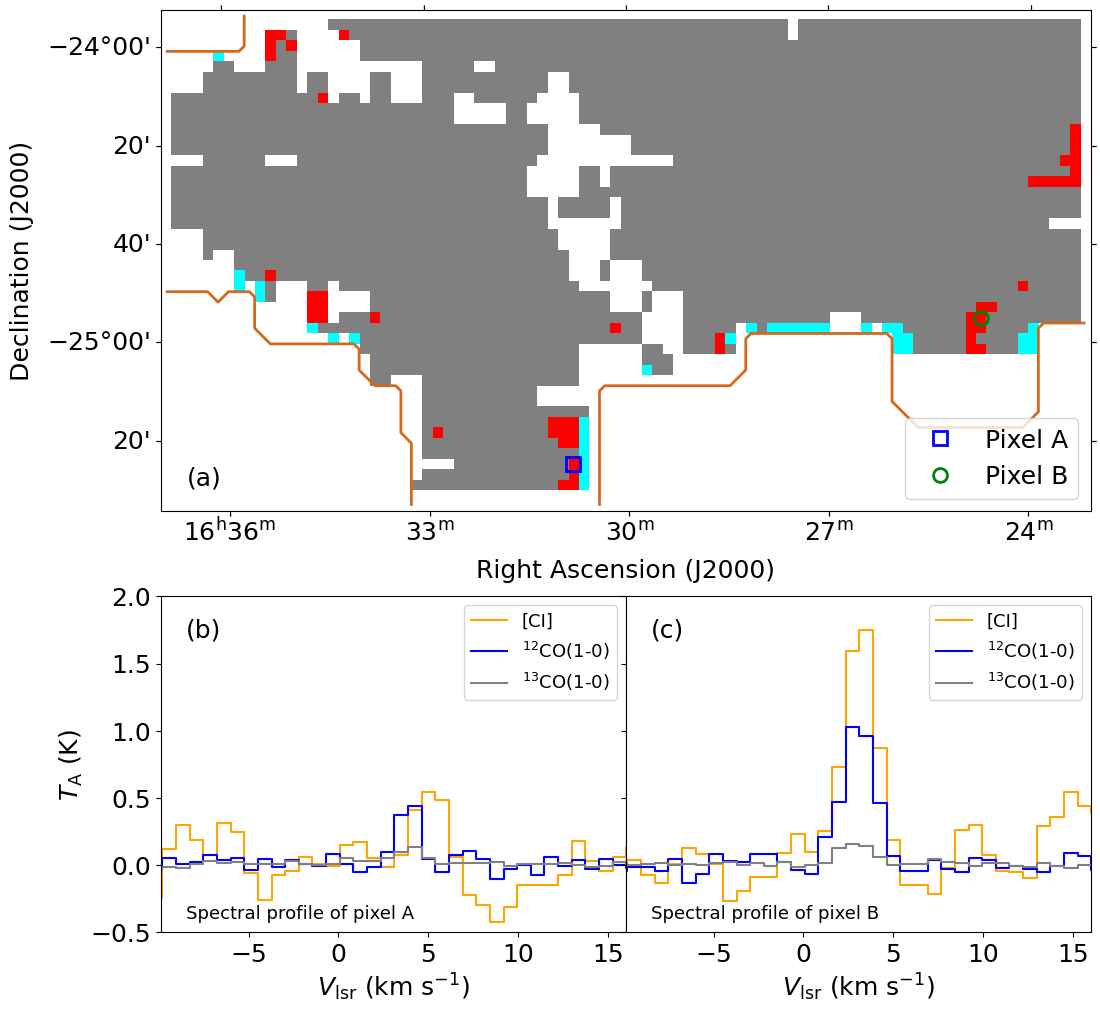}
 \caption{(a): Spatial distribution of different mask regions based on \ci\ and CO(1-0) emission. The edge of CO data is defined by a solid chocolate-colored line. Regions where \ci\ intensity is weaker than CO(1-0) intensity (W(\ci) $<$ W(CO(1-0))) are shown in gray, while stronger \ci\ regions (W(\ci) $>$ W(CO(1-0))) are shown in red and cyan. Among the latter, the cyan pixels represent areas that would be excluded by applying a two-pixel buffer beyond the data boundary, a threshold implemented to mitigate convolution artifacts from external null values.  (b) and (c): The spectra of two representative positions, A and B,  where the \ci\ emission is stronger than that of \co. }
 \label{fig:CO_dark}
 \end{figure*}

 As shown in Fig.~\ref{fig:CO_dark}, the integrated intensity of \ci\ exceeds that of CO(1-0) in about 2.0\% of the pixels near the molecular cloud edges. To mitigate convolution effects at the CO boundary, pixels within one beam width of the edge were trimmed prior to analysis. This provides direct evidence that \ci\ effectively traces CO-dark molecular gas. More details will be discussed in the Section \ref{subsec:co-dark_discussion}.

\subsection{Excitation Temperature and Column Density of \ci\ and CO}

Since the critical density of \ci\ and $^{12}$CO(1-0) %TB: say what is the critical density of CO(1-0) here too
lines are comparable, similar excitation conditions of these lines are expected \cite{2002ApJS..139..467I}.
We assume that the excitation temperature T$\rm_{ex}$ of \ci\, is equal to that of $^{12}$CO. When the emission is optically thick, T$\rm_{ex}$ can be written as

\begin{equation}
T_{ex} = \frac{5.53}{ln(1+5.53/[T_B(^{12}\rm{CO})+0.819])} \ K,
\end{equation}
where $T\rm_B(^{12}CO)$ is the brightness temperature of $^{12}$CO(1-0). The cosmic background temperature is taken to be 2.7~K. As presented in Fig. \ref{fig:spectra}, there exists significant self-absorption feature in \co\ spectra toward dense regions. We follow the analysis in Taurus region \cite{2008ApJ...680..428G} and carry out a self-absorption analysis to recover the true \co\ spectrum. A maximum $T\rm_{ex}$ value of 48 K was found around Peak 2. The value of $T\rm_{ex}$ ranges from 2.81 K to 48.4 K, with L1688 showing significantly higher $T\rm_{ex}$. The maximum $T\rm_{ex}$ in L1688 is 48.4 K, while in L1689 it is 30.74 K.

After determining $T\rm_{ex}$, the column densities of \ci\ and CO are derived under the assumption of local thermodynamic equilibrium (LTE). 
The critical densities of $^{12}$CO and \ci\ are comparable, supporting the assumption that the $^{12}$CO (J=1-0) and \ci\ lines share similar excitation conditions \cite{2021PASJ...73..174I}. Although additional mechanisms—such as the recombination of C$^+$ with electrons—can contribute to the excitation of \ci\, their overall impact is limited. As discussed by \cite{2012ApJS..203...13G} and references therein, excitation by electron collisions is constrained by both the available electron abundance and the total gas density. In the $\rho$ Oph molecular cloud, the \h2\ 
density is generally high; thus, even if all carbon were ionized, collisions with \h2\ would still occur at rates four orders of magnitude higher than those with electrons. Although the electron–\ci\ collisional cross section is roughly two orders of magnitude larger than that for \h2, this enhancement is insufficient to compensate for the much lower electron density. Consequently, the excitation of \ci\ remains close to local thermodynamic equilibrium (LTE), and the contribution from electron collisions is negligible. Under LTE assumption,  the column density of \ci , N(\ci), is calculated through  \cite{1981ApJ...251..533P,2021PASJ...73..174I} :
\begin{equation}
N(\textrm{[C\,{\sc i}]}) = 1.98\times 10^{15}e^{\frac{E_1}{kT_{ex}}}f_{\tau}f_{T_{ex}}Q(T_{ex})\int T_{b}(\textrm{[C\,{\sc i}]})d\upsilon\ cm^{-2} ,
\end{equation}
 where $T\rm_{ex}$, $\tau$ and $T\rm_b$(\ci)  denote the excitation temperature,  optical depth, and brightness temperature of the \ci\ 492 GHz transition, respectively. The correction factor for \ci\ optical depth and excitation temperature are given by  $f_{\tau}= \tau/(1-e^{-\tau})$ and  $f_{T_{ex}}=(1-J_{\nu}(T_{bg})/J_{\nu}(T_{ex}))^{-1}$, where the background temperature is T$_{bg}$ $= 2.73$ K. The radiation function $J_{\nu}$ and partition function $Q(T_{ex})$ can be expressed with $J_{\nu}= \frac{h\nu/k}{exp(h\nu/kT)-1}$ and $Q(T_{ex})= 1+3exp(-E_1/kT_{ex})+5exp(-E_2/kT_{ex})$ , where $E_1$\,=\,23.5\,K and $E_2$\,=\,62.5\,K represent the energies of  $^3$P$_1$$\rightarrow$$^3$P$_0$ and $^3$P$_2$ $\rightarrow$$^3$P$_1$ transitions, respectively.

\begin{figure*}[ht!]
\centering
 \includegraphics[width=0.98\textwidth]{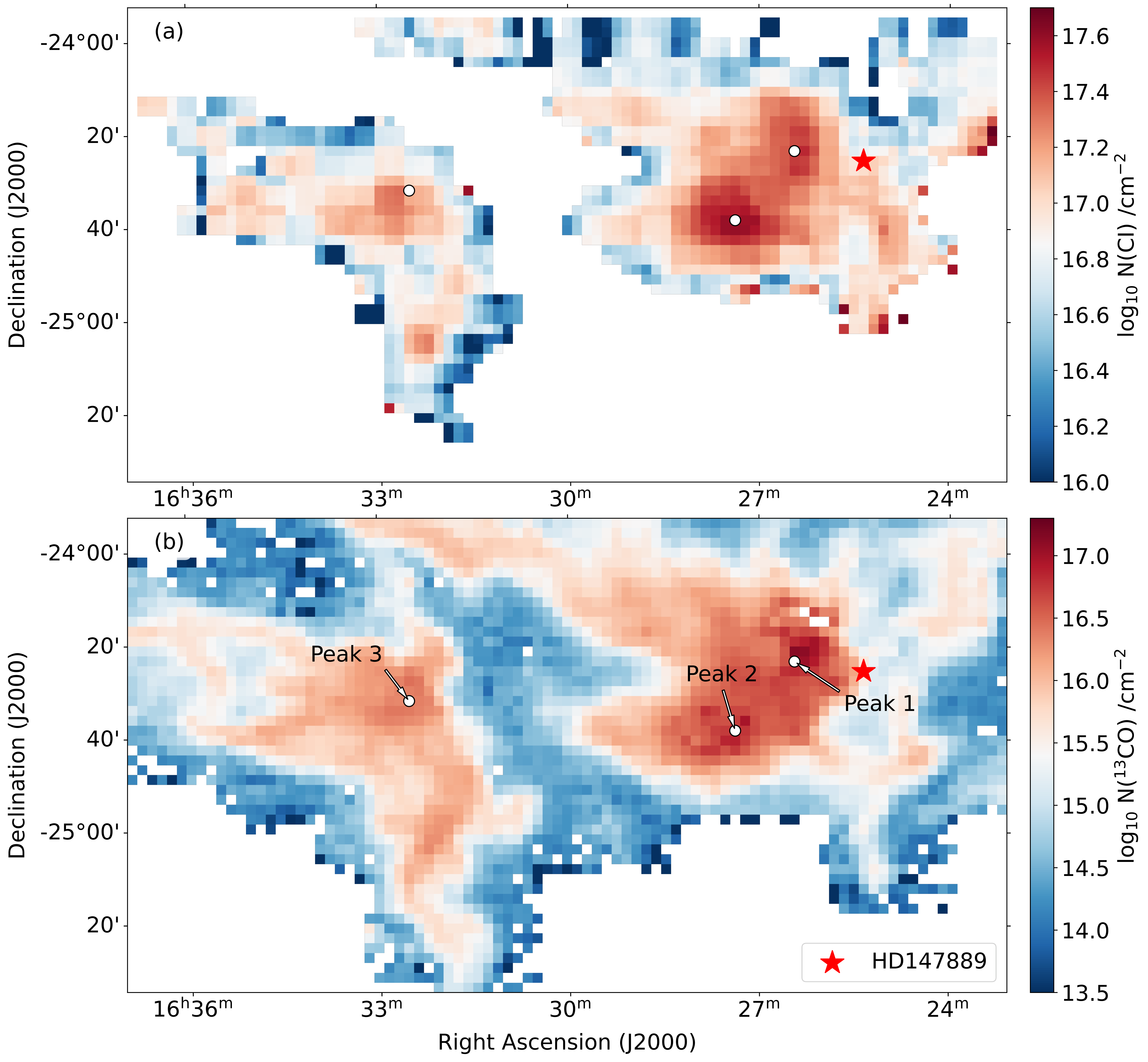}
 \caption{Spatial distribution of column densities N(\ci) (Panel a) and N($^{13}$CO) (Panel b). The three peak of W(\ci) are marked on the N($^{13}$CO) map. }
 \label{fig:N}
 \end{figure*}

As shown in Fig. \ref{fig:N}(a), $N$(\ci) ranges from 4.85 $\times$ 10$^{14}$ cm$^{-2}$ to 6.29 $\times$ 10$^{17}$ cm$^{-2}$, with an average value of (9.00 ${\pm}$ 0.17) ${\times}$ 10$^{16}$ cm$^{-2}$ in the \r\ Oph region.
The $N$(\ci) value in L1688 is higher than that in L1689, as the average \ci\ column density is (1.06${\pm}$ 0.2) $\times$ 10$^{17}$ cm$^{-2}$, while it is  (6.8${\pm}$ 0.16) $\times$ 10$^{16}$ cm$^{-2}$ in L1689. 
The maximum value of $N$(\ci) in L1688 is 6.29 $\times$ 10$^{17}$ cm$^{-2}$ while it is 3.84 $\times$ 10$^{17}$ cm$^{-2}$ in L1689. The minimum value of $N$(\ci) in L1688 is 4.85 $\times$ 10$^{14}$ cm$^{-2}$ while in L1689, it is 7.34 $\times$ 10$^{14}$ cm$^{-2}$. 

The optical depth of $^{13}$CO( 1-0), $\tau_{13}$ can be obtained by assuming the same excitation temperature of $^{12}$CO( 1-0), thus 
% \begin{equation}
% \tau_{13} = -ln\big[1- \frac{T_B^{13}}{5.29[J_1(T_{ex})-0.164)]}\big] \,
% \end{equation}

\begin{equation}
\tau_{13} = -\ln\bigg[1- \frac{\text{  $T_B(^{13}CO)$}}{5.29[J_1(T_{ex})-0.164)]}\bigg] \,,
\end{equation}
in which $J_1(T_{ex})=1/[exp(5.29\,\mathrm{K}/T_{ex})-1]$. $T_B(^{13}CO)$ represents the brightness temperature of $^{13}$CO(1-0). The column density of $^{13}$CO can be written as: 

\begin{equation}
    N(^{13}CO)=2.42 \times 10^{14}\dfrac{ \displaystyle\int \text{$T_B(^{13}CO)$} \,dv}{1-e^{-5.29/T_{ex}}} \times \dfrac{\tau_{13}}{1-e^{-\tau_{13}}} \ \text{cm}^{-2}.
\end{equation}

As shown in Fig. \ref{fig:N}(b), the average $^{13}$CO column density in L1688 is (7.93${\pm}$ 0.37) $\times$ 10$^{15}$ cm$^{-2}$ while in L1689, it is (5.98${\pm}$ 0.12) $\times$ 10$^{15}$ cm$^{-2}$. The average value in the $\rho$ Oph region is (5.62${\pm}$ 0.19)$\times$ 10$^{15}$ cm$^{-2}$. The minimum value in L1688 is 2.45 $\times$ 10$^{14}$ cm$^{-2}$ and in L1689 is 3.62$\times$ 10$^{14}$ cm$^{-2}$. The maximum value in L1688 is 1.34 $\times$ 10$^{17}$ cm$^{-2}$ and  is 2.50$\times$ 10$^{16}$ cm$^{-2}$ in L1689. 

By adopting a $\rm ^{12}CO/^{13}CO$ ratio of 70 \cite{1994ARA&A..32..191W}, N(\13co) is converted into N(\co) through $N$(\co)=$70 \times N$(\13co).

The column density values of \co\  near Peak 1 and Peak 2 (see Fig. \ref{fig:cicontour}) are ($3.10 \pm 0.15) \times 10^{17}$ cm$^{-2}$ and ($3.84 \pm 0.19) \times 10^{17}$ cm$^{-2}$, respectively, which are consistent with but slightly lower than the values of $5.50 \times 10^{17}$ cm$^{-2}$ and $5.00 \times 10^{17}$ cm$^{-2}$ reported by \cite{2003ApJ...589..378K}.

\subsection{Physical and Chemical Properties of the $\rho$ Oph Cloud}
\label{subsec:phy_chem}

\subsubsection{UV Intensity Based on $DUSTY$ Model}
\label{subsubsec:dusty}

The external UV radiation field acts as the primary heating source and plays a critical role in regulating molecular formation and destruction within molecular clouds. Consequently, it significantly influences the carbon cycle, particularly the transitions between [C\,{\sc ii}], \ci\, and CO. 

In our previous study \cite{2022RAA....22h5017X}, we developed a method to estimate the UV radiation intensity in nearby molecular clouds of the Gould Belt, including the $\rho$ Ophiuchi region. This approach leverages radiative transfer modeling of dust employing the \texttt{DUSTY} code\cite{1997MNRAS.287..799I}, utilizing dust temperature and $N$(H$_2$) maps derived from the Herschel Gould Belt Survey. Specifically, we performed a series of \texttt{DUSTY} simulations assuming a plane‑parallel slab geometry with perpendicular illumination by an external UV field. The model takes as inputs the dust properties, the slab optical depth, and the incident UV spectrum, and returns the dust temperature as a function of optical depth. We then used the observed and H$_2$ column‑density maps (the latter converted to optical depth via the standard extinction relation) to compare with the model grids, thereby deriving the UV intensity at each pixel.
The analysis assumed a plane-parallel slab geometry with direct illumination by FUV photons. However, this simplified treatment may introduce uncertainties, as the incident angle between the UV radiation field and the slab plane can vary spatially. Despite these limitations, the method provides a quantitative framework for assessing the UV contribution across different molecular regions. The G$_0$ map is derived from dust radiative transfer modeling using the T$_{dust}$ and N(H$_2$) data from HGBS.

\begin{figure*}[ht!]
\centering 
\includegraphics[width=0.98\textwidth]{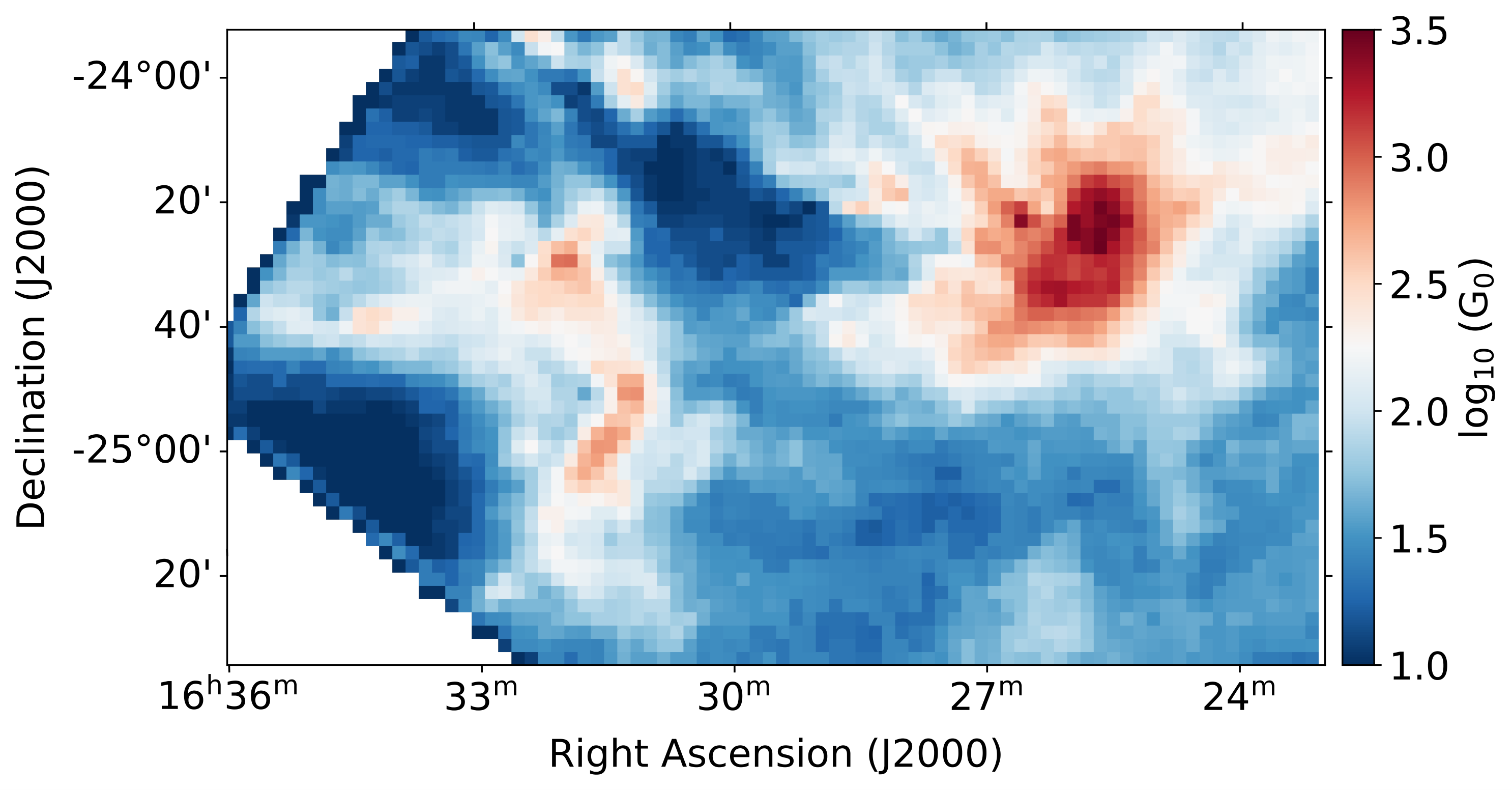}
\caption{The spatial distribution of UV intensity ($G_0$) in $\rho$ Oph region.} 
\label{fig:G0_map}
\end{figure*}

Fig. \ref{fig:G0_map} presents the spatial distribution of the UV radiation field. The FUV intensity ($G_0$, in units of standard Habing field of $1.6\times 10^{-6}$ W m$^{-2}$) peaks near the massive star HD 147889, reaching values of $\sim3\times10^3$. The mean $G_0$ value is (161 $\pm$ 4.16) in L1688 and (68 $\pm$ 2.11) in L1689.

\subsubsection{Species Abundances}
\label{subsubsec:abundances}

\begin{figure*}[ht!]
\centering
 \includegraphics[width=0.98\textwidth]{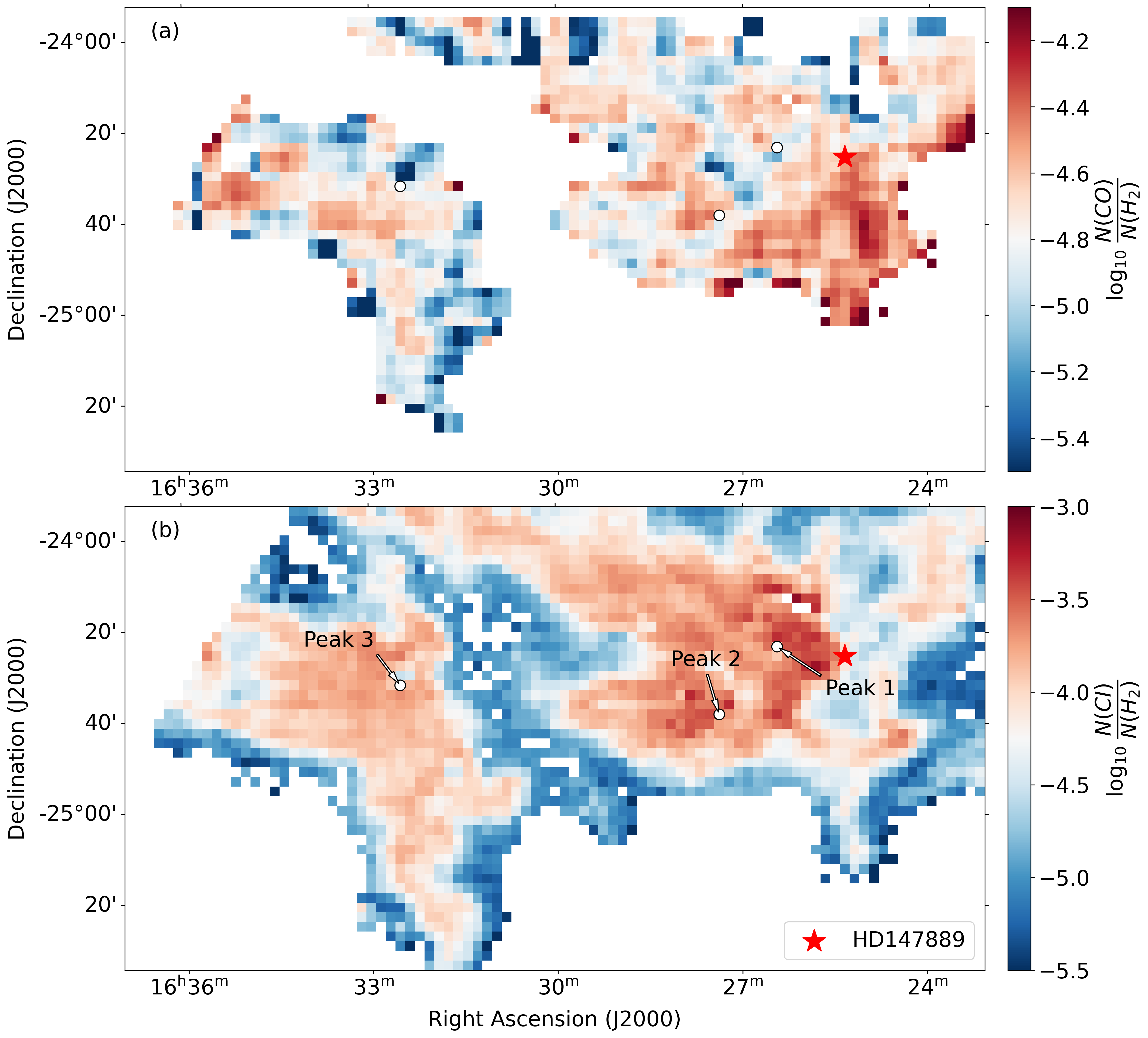}
 \caption{Abundance distributions of \ci\ (Panel a) and $^{12}$CO (Panel b) across the $\rho$ Oph molecular cloud, shown on a logarithmic scale. The locations of HD 147889 (red star) and Peaks 1–3 (solid dots) are marked.}
 \label{fig:xci_xco}
 \end{figure*}

The abundance of a species relative to H$_2$ is defined as $f_{\rm X}$ = $N_{\rm X}$/$N_{\rm H_2}$. We investigate the spatial distributions of \ci\ and CO abundance (Fig.~\ref{fig:xci_xco}(a) and Fig.~\ref{fig:xci_xco}(b)), examining their dependence on \h2\ column density N(\h2) (in units of cm$^{-2}$) and UV radiation field G$_0$ (in units of standard Habing field). 

\begin{figure*}[ht!]
\centering
  \includegraphics[width=0.7\textwidth]{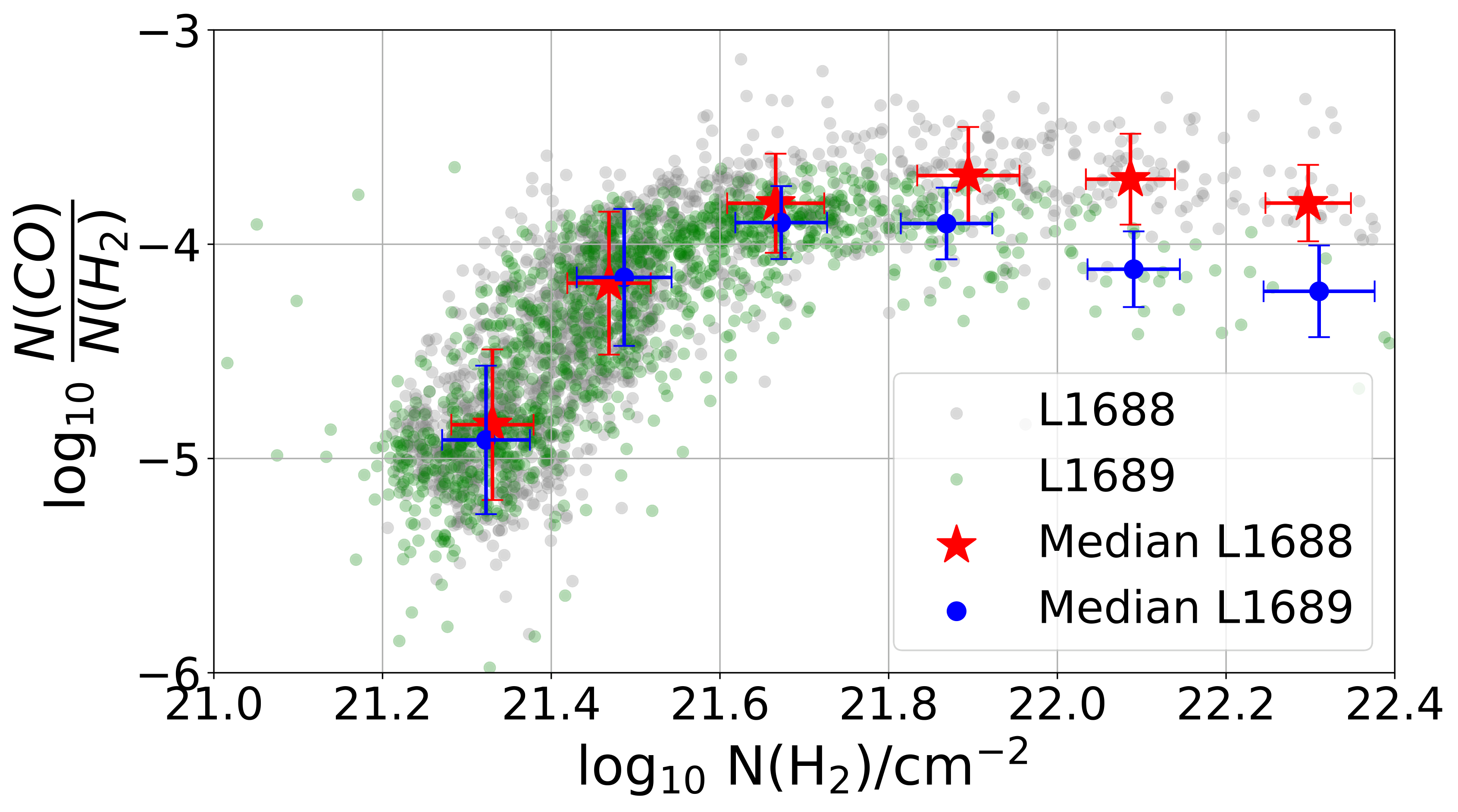}
\caption{Scatter plot showing the relationship between $\log_{10} (N(\text{CO})/N(\text{H}_2))$ and $N(\text{H}_2)$. Data are divided into two molecular cloud regions: L1688 (gray points) and L1689 (green points). Median values were computed by binning the data in $\log_{10} N(\text{H}_2)$ from 21.2 to 22.4 with 0.2 dex intervals. For each bin, we calculated the median of both X (hydrogen column density) and Y (CO abundance) values. Error bars represent the statistical uncertainties. Median values are shown as red stars (L1688) and blue squares (L1689) with connecting lines.}
 \label{fig:co_abundance_h2}
 \end{figure*}

As shown in Fig.~\ref{fig:xci_xco}(b), the abundances of $^{12}$CO are in the range of 7.56 $\times$ 10$^{-6}$ to 6.42 $\times$ 10$^{-4}$ in L1688, and 8.33 $\times$ 10$^{-6}$ to 2.49 $\times$ 10$^{-4}$ in L1689. The mean abundances of $^{12}$CO are (7.61$\pm$ 0.29) $\times$ 10$^{-5}$ and (9.23$\pm$ 0.22)  $\times$ 10$^{-5}$ toward L1688 and L1689, respectively. 
We find that CO abundance peaks in the interior of the clouds and is lowest at the edges. The CO abundance at the east portion of HD\,147889 decreases by an order of magnitude, possibly due to  dissociated by the FUV photons emitted from massive star. 
In order to present a statistical distribution, the dependence of CO abundance on H$_2$ column density ($N_{\rm H_2}$) is shown in Fig.~\ref{fig:co_abundance_h2}. The median value of CO abundance increases by almost an order of magnitude from ${\rm log_{10}} \ N_{\rm H_2} \approx 21.2$ to ${\rm log_{10}} \ N_{\rm H_2} \approx 21.6$. The increasing trend turns to flatten for L1688 when ${\rm log_{10}} \ N_{\rm H_2} > 21.6$ (corresponding to $A_{\rm V}$ of 4\,mag). This trend is reasonably consistent with that of Taurus cloud \cite{2010ApJ...721..686P} and  IC~348 in Perseus molecular cloud at similar extinction range and FUV intensity \cite{2023ApJ...942..101L}. 
The abundance of CO slightly decreases at higher $N_{\rm H_2}$ (${\rm log_{10}} \ N_{\rm H_2} > 21.6$), which may be related to CO depletion at high column densities \cite{1998ApJ...499..234C}.

Compared to that of \co,  the spatial distribution of \ci\ abundance does not show a clear enhancement in the interior of the clouds, but it is especially higher around the massive star (Fig.~\ref{fig:xci_xco}(a)). The abundance of \ci\ ranges from 2.24 $\times$ 10$^{-7}$ to 2.39 $\times$ 10$^{-4}$ at L1688, with a mean value of (1.99$\pm$ 0.06) $\times$ 10$^{-5}$. In L1689, the abundance of \ci\ ranges from 4.43 $\times$ 10$^{-7}$ to 1.06 $\times$ 10$^{-4}$, with a mean value of (1.40$\pm$ 0.13) $\times$ 10$^{-5}$.

\begin{figure*}[ht!]
\centering
 \includegraphics[width=0.96\textwidth]{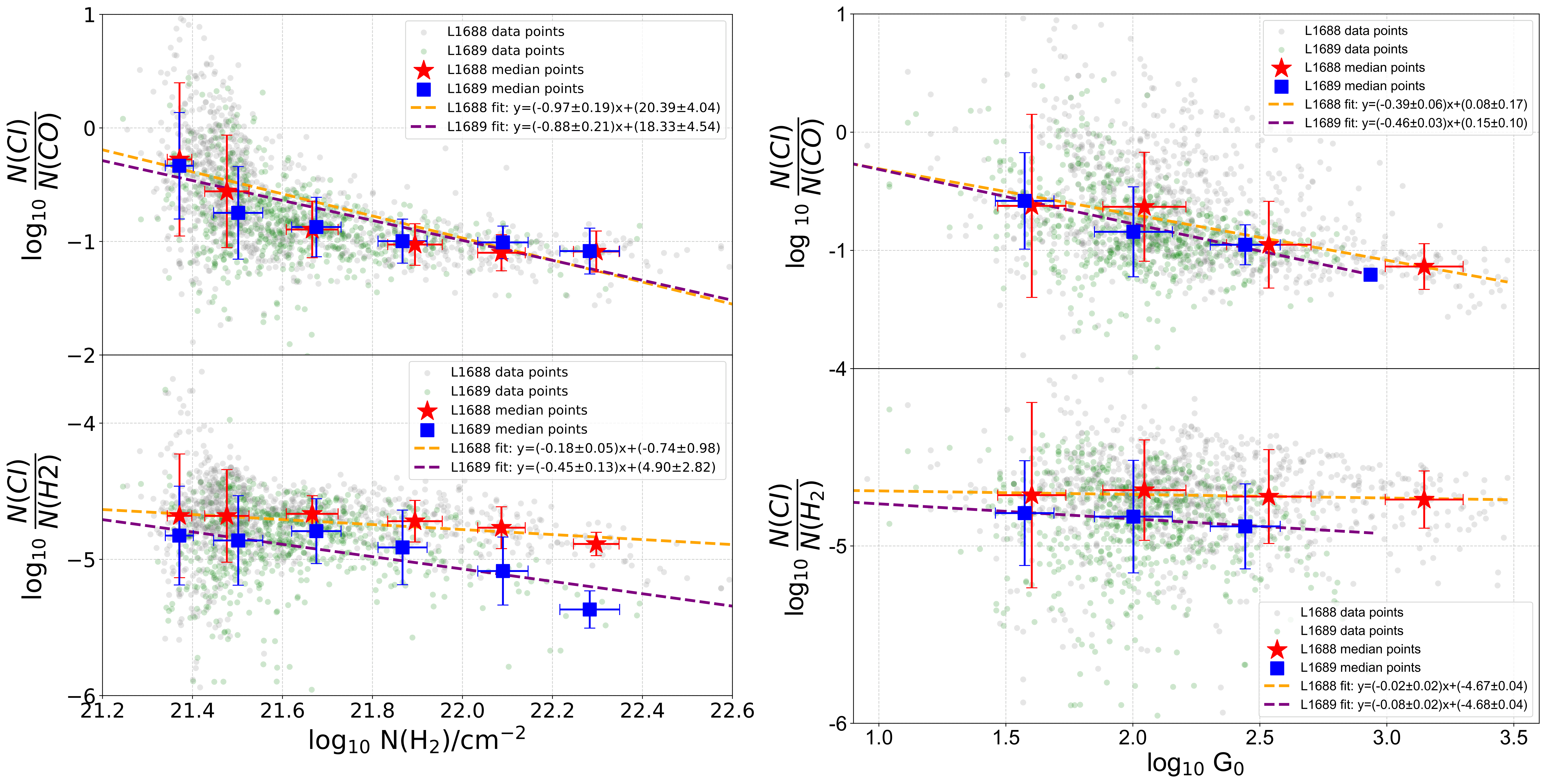}
\caption{The left panel shows scatter plots of the relationship between $N({CI})/N({CO})$(upper), $N({CI})/N({H_2})$(lower), and UV radiation intensity. Data are separated into two molecular cloud regions: L1688 (gray points) and L1689 (green points). Median values were computed by binning the $\log_{10}$ G$_0$ radiation intensity from 1.1 to 3.5 with 0.6\,dex intervals. For each bin, we calculated the median of both $X$ ($\log_{10}$ G$_0$) and $Y$ ($N({CI})/N({H_2})$ or $N({CI})/N({CO})$ ratio) values, with error bars representing statistical uncertainties. These median values are displayed as red stars (L1688) and blue squares (L1689). The right panel follows a similar plotting methodology as Fig\,\ref{fig:co_abundance_h2}. The fitting formula obtained through the median is displayed in the legend, with specific numerical values also presented in Table\ref{tab:fitting_results}.}

 \label{fig:xci_G0_NH2}
 \end{figure*}

The dependence of N(\ci)/N(\h2) and N(\ci)/N(CO) on $G_0$ and $N_{\rm H_2}$ is shown in Fig.~\ref{fig:xci_G0_NH2}. To better investigate this variation trend, we performed linear regression analysis to investigate the relationships between atomic carbon abundance ratios and environmental parameters. The results of linear fitting   are summarized in Table~\ref{tab:fitting_results}.

The N(\ci)/N(CO) ratio shows a strong inverse correlation with $G_0$ in both regions. In contrast, the N(\ci)/N(H$_2$) ratio decreases more gradually with increasing $G_0$, reflecting a weaker but still negative dependence on the UV radiation field.

The N(\ci)/N(CO) value decreases from 0.5 at log$_{10}$N(\h2)=21.4 to 0.1 at log$_{10}$N(\h2)=21.9. It remains at a almost constant value of $\sim 0.1$ when log$_{10}$N(\h2)$>$21.9.  This relationship is consistent with [\ci] studies on massive star-forming regions: Orion A and Orion B \cite{2002ApJS..139..467I} and RCW 38 \cite{2021PASJ...73..174I}.  

The N(\ci)/N(\h2) ratio exhibits a significantly stronger negative correlation with N(\h2) in L1689 than in L1688. However,  in contrast to the behavior of the N(CO)/N(\h2) ratio in Fig. \ref{fig:co_abundance_h2}, the N(\ci)/N(\h2) ratio remains nearly constant for N(\h2) when ${\rm log_{10}} \ N_{\rm H_2} < 21.9$. This weak dependence on N(\h2) indicates that \ci\ serves as a more reliable tracer of \h2\ than CO in diffuse and translucent regions.

We suggest that the \ci\,-to-CO transition occurs at $21.4 < {\rm log_{10}} \ N_{\rm H_2} < 21.6$, above which the abundance of CO gradually increases to its canonical value. 

\begin{table*}[htbp]
	\centering
	\caption{Linear fitting results for logarithmic abundance ratios with uncertainties.}
	\label{tab:fitting_results}
	\small
	\begin{tabular}{lcccccccc}
		\toprule
		& \multicolumn{2}{c}{$\log_{10}[\mathrm{CI}/\mathrm{CO}]$ vs $\log_{10} N(\mathrm{H}_2)$} 
		& \multicolumn{2}{c}{$\log_{10}[\mathrm{CI}/\mathrm{CO}]$ vs $\log_{10} G_0$} 
		& \multicolumn{2}{c}{$\log_{10}[\mathrm{CI}/\mathrm{H}_2]$ vs $\log_{10} N(\mathrm{H}_2)$} 
		& \multicolumn{2}{c}{$\log_{10}[\mathrm{CI}/\mathrm{H}_2]$ vs $\log_{10} G_0$} \\
		\cmidrule(lr){2-3} \cmidrule(lr){4-5} \cmidrule(lr){6-7} \cmidrule(lr){8-9}
		Parameter & L1688 & L1689 & L1688 & L1689 & L1688 & L1689 & L1688 & L1689 \\
		\midrule
		A & $-0.97 \pm 0.19$ & $-0.88 \pm 0.21$ & $-0.39 \pm 0.06$ & $-0.46 \pm 0.03$ & $-0.18 \pm 0.05$ & $-0.45 \pm 0.13$ & $-0.02 \pm 0.02$ & $-0.08 \pm 0.02$ \\
		B & $20.39 \pm 4.04$ & $18.33 \pm 4.54$ & $0.08 \pm 0.17$ & $0.15 \pm 0.10$ & $-0.74 \pm 0.98$ & $4.90 \pm 2.82$ & $-4.67 \pm 0.04$ & $-4.68 \pm 0.04$ \\
		\bottomrule
	\end{tabular}
	\vspace{0.2cm}
	
	\small
	\raggedright
	Note: All fitting results are based on the linear form $Y = A \times X + B$, where $Y$ represents either $\log_{10}[N(\mathrm{CI})/N(\mathrm{CO})]$ or $\log_{10}[N(\mathrm{CI})/N(\mathrm{H}_2)]$, and $X$ represents either $\log_{10} N(\mathrm{H}_2)$ or $\log_{10} G_0$.
\end{table*}

\section{PDR Modeling}
\label{pdr_model}

For the purposes of further comparison with theoretical models, we use the publicly available code {\sc 3d-pdr} \cite{Bisbas12}, which treats astrochemistry of any density distribution in 1D and 3D. By taking into consideration several heating and cooling processes, the code performs iterations over thermal balance and terminates once the total heating is approximately equal to the total cooling. We use a subset of the UMIST2012 chemical network \cite{McElroy13} consisting of 33 species and 330 reactions. Here, we only consider the elements of carbon, oxygen, helium and hydrogen in the gas-phase. The initial elemental abundances used are $\rm C/H=1.4\times10^{-4}$ \cite{Cardelli96,Draine11} and $\rm O/H=3\times10^{-4}$ \cite{Cartledge04,Draine11}, normalized to total hydrogen. These abundances are lower than the Solar abundances reported in \cite{Asplund09} due to depletion on grains. The dust-to-gas mass ratio used is set to $10^{-2}$ at all times.  The formation of H$_2$ follows the treatment of \cite{Cazaux02,Cazaux04,Cazaux10}. We evolve the chemistry for 10~Myr, by which time it has reached chemical equilibrium \cite{Bayet09,Holdship22}. We perform two suites of models to explore the trends concerning the value of the cosmic-ray ionization rate, and the trends concerning the relation between the FUV intensity and the local number density.

%Cosmic-rays results:
In the first suite, we adopt a density distribution based on the $A_{\rm V,eff}-n_{\rm H}$ relation presented in \cite{Bisbas23} (see also \cite{2025A&A...700L..16G}). This one-dimensional density slab is found to reproduce reasonably well the PDR properties of the more complex three-dimensional density distributions at a minimal computational cost. Therefore, we use this distribution and we perform PDR modelling by varying the FUV radiation field in the range $\chi/\chi_0=10^0-10^4$ and the cosmic-ray ionization rates in the range $\zeta_{\rm CR}=10^{-17}-10^{-14}\,{\rm s}^{-1}$. With a 0.1~dex precision in these values, we perform a total of $\sim1.7\times10^3$ PDR models. We then calculate the column densities of N(H$_2$), N(C{\sc i}) and N(CO) and compare their mutual ratios with those derived from observations.

\begin{figure*}
    \centering
    \includegraphics[width=0.92\linewidth]{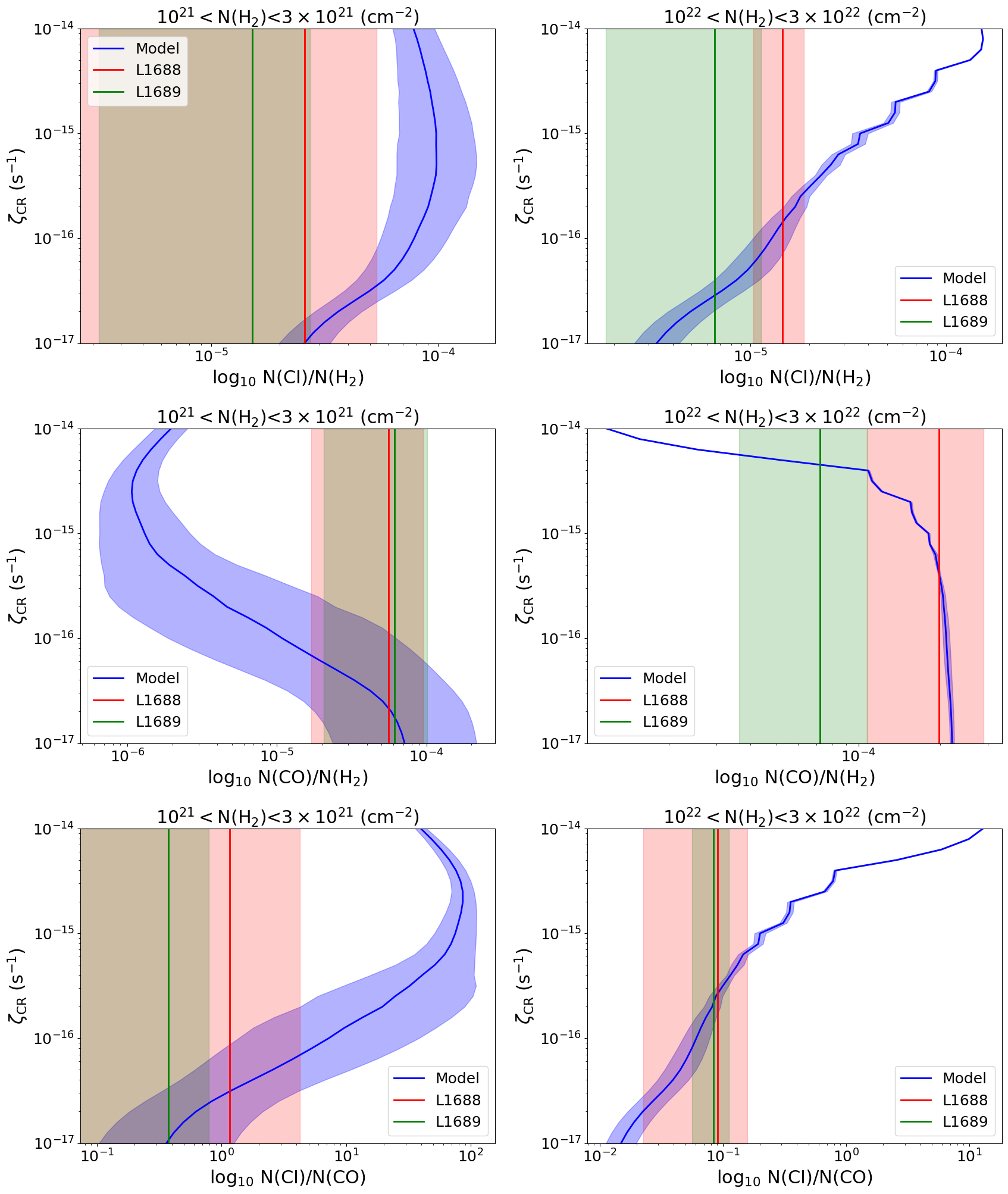}
    \caption{Results from the {\sc 3d-pdr} modeling showing relationships between N(C{\sc i})/N(H$_2$) (top row), N(CO)/N(H$_2$) (middle row), N(C{\sc i})/N(CO) (bottom row) and $\zeta_{\rm CR}$. Left column corresponds to $\rm 10^{21}<N(H_2)<3\times10^{21}\,{\rm cm}^{-2}$ and right column  to $\rm 10^{22}<N(H_2)<3\times10^{22}\,{\rm cm}^{-2}$. The blue stripe corresponds to PDR models using the variable $A_{\rm V,eff}-n_{\rm H}$ slab while the red and green stripes to L1688 and L1689 observations, respectively. The width represent $1\sigma$ standard deviation of the mean due to the influence of $G_0$. In general we find that in low H$_2$ column densities the cosmic-ray ionization rate is expected to be low and close to the mean galactic value, while higher H$_2$ column densities show an enhanced value of $\zeta_{\rm CR}$, likely attributed to the on-going star formation.}
    \label{fig:CRresults}
\end{figure*}

Fig.~\ref{fig:CRresults} illustrates our results for the first suite of models. The stripes correspond to the the average values of the aforementioned PDR models (in blue color) and the two regions (L1688 in red, L1689 in green) while the shaded region represents the $1\sigma$ standard deviation around the mean due to the different G$_0$ intensities. In regions of low N(H$_2$) (left column), we find that the cosmic-ray ionization rate is low ($\zeta_{\rm CR}\sim1-3\times10^{-17}$) with an upper limit of $\sim10^{-16}\,{\rm s}^{-1}$, very similar to the average Galactic value \cite{Dalgarno06,Gaches22,Obolentseva24}. In regions of high N(H$_2$) (right column), we observe an interesting feature. Although from the top right (N(\ci)/N(H$_2$)) and bottom right (N(\ci)/N(CO)) panels we are able to identify solutions for the value of $\zeta_{\rm CR}$, the N(CO)/N(H$_2$) ratio (middle right panel) shows high degeneracy for the L1688 region and a very high $\zeta_{\rm CR}$ value for the L1689 region. It is only when accounting for the \ci\ column density that we are able to find reasonable solutions for $\zeta_{\rm CR}$, strongly supporting the ability of \ci\ line to break degeneracies. Using this line, we can identify a $\zeta_{\rm CR}\sim2-10\times10^{-16}\,{\rm s}^{-1}$ for the L1688, and a $\zeta_{\rm CR}\sim2\times10^{-17}-8\times10^{-16}\,{\rm s}^{-1}$ for the L1689. A trend of an increasing $\zeta_{\rm CR}$ with N(H$_2$) is likely to be connected with more intense conditions due to ongoing star-formation \cite{Gaches19}.

\begin{figure*}
    \centering
    \includegraphics[width=0.55\linewidth]{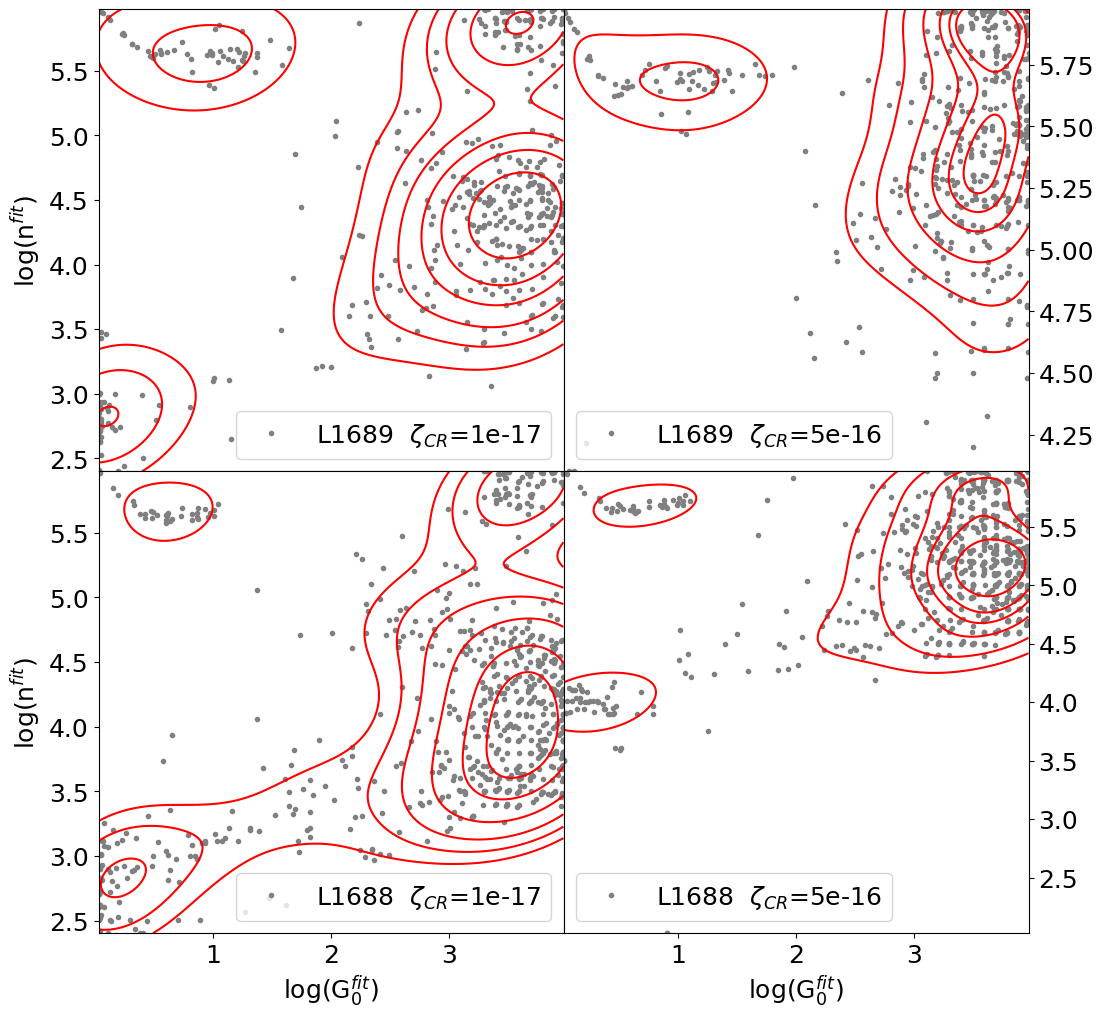}
    \caption{MCMC results of L1689 (top panels) and L1688 region (bottom panels). The density contours of fitted points are marked with red color. The x-axis shows the logarithm of the G$_0$, while the y-axes (both left and right) represent the logarithm of the fitted volume density. }
    \label{fig:mcmc_results}
\end{figure*}

%MCMC results: 
In the second suite of models, we construct a set of uniform density clouds with densities in the range $n_{\rm H}=10^2-10^6\,{\rm cm}^{-3}$ at 0.1~dex precision. We use the above FUV intensity range and we consider two cosmic-ray ionization rates, based on the aforementioned results. A low one of $\zeta_{\rm CR}=10^{-17}\,{\rm s}^{-1}$ and a higher one of $\zeta_{\rm CR}=5\times10^{-16}\,{\rm s}^{-1}$. We then calculate the column densities of H$_2$, \ci\ and CO at every depth point in the models and perform  Markov chain Monte Carlo (MCMC) analysis to explore the trends connecting the local density and the intensity of the FUV radiation field. Fig. \ref{fig:mcmc_results} shows the results of our calculations. For both $\zeta_{\rm CR}$ values and in both regions, we find that the FUV intensity $G_0$ increases with increasing density $n_{\rm H}$. Eventually, we identify two possible solutions; one for a low $n_{\rm H}-G_0$ pair of values and one for a higher $n_{\rm H}-G_0$ pair of values. The best-fit number density in both these solutions is found to be strongly depended on the value of the cosmic-ray ionization rate. In particular, for both clouds and for $\zeta_{\rm CR}=10^{-17}\,{\rm s}^{-1}$ we find $10^3\lesssim n_{\rm H}\lesssim3\times10^3\,{\rm cm}^{-3}$ and $3\times10^3\lesssim n_{\rm H}\lesssim10^4\, {\rm cm}^{-3}$ for $G_0\lesssim10$ and $G_0\gtrsim10^3$, respectively. Considering the higher value of $\zeta_{\rm CR}=5\times10^{-16}\,{\rm s}^{-1}$, the best-fit densities are shifted upwards by approximately an order of magnitude i.e. $10^4\lesssim n_{\rm H}\lesssim3\times10^4\, {\rm cm}^{-3}$ and $3\times10^4\lesssim n_{\rm H}\lesssim3\times10^5\, {\rm cm}^{-3}$ for $G_0\lesssim10$ and $G_0\gtrsim10^3$, respectively. As we do not know unambiguously the $\zeta_{\rm CR}$ value in these clouds, we are not able to provide distinct solutions for the $n_{\rm H}-G_0$ pairs.  However, it is clear that the ISM conditions in the collapsing regions of both clouds are much different than their surrounding medium with both $G_0$ and $\zeta_{\rm CR}$ to increase.

\section{Discussion}\label{sec:discussion}

\subsection{Is \ci\ a better molecular gas tracer than CO?}
\label{subsec:co-dark_discussion}

As shown in Section \ref{subsec:intg-codark} and \ref{subsec:phy_chem}, we found that \ci\ traces hydrogen molecules better in the region of stronger UV radiation field compared to CO.
The CO/H$_2$ ratio exhibits a much stronger dependence on the UV radiation field compared to the \ci/H$_2$ ratio. However, CO is a more stable tracer than \ci\ in the region of the higher column density of hydrogen molecules.

CO-dark gas, which is primarily composed of molecular hydrogen ($H_{2}$)  \cite{2010ApJ...716.1191W}, plays a crucial role in the star formation processes and overall evolution of galaxies. The low  opacity of CO-dark gas to CO emissions presents a challenge in directly tracing this elusive component of the interstellar medium\cite{2005Sci...307.1292G}.

However, by observing the \ci\ line, we can gain valuable insights into the presence and distribution of CO-dark gas \cite{2010ApJ...716.1191W,2015MNRAS.448.1607G}.
The \ci\ line serves as an alternative tracer that can penetrate the regions where CO is dissociated due to the influx of far-ultraviolet (FUV) radiation. 
This is particularly relevant in low-metallicity environments, where the dust content is significantly reduced \cite{1999ApJ...513..275B}, allowing FUV photons to penetrate deeper into the molecular clouds, leading to the PDR regions.
Studies have shown that \ci\ can provide a more direct conversion from observed line luminosities to molecular gas masses, offering a clearer picture of the total gas reservoir available for star formation\cite{{2020A&A...643A.141M}}. Our study of $\rho$ Oph molecular cloud support this point (Section \ref{subsubsec:abundances}). This conversion has been successfully applied in various galaxies, including those with low-metallicity environments, where traditional CO-based methods fail to account for the significant amount of dark gas. 

As shown in Section \ref{subsec:intg-codark}, the \ci\ intensity exceeds CO intensity toward 2.0\% pixels, indicating the existence of CO-dark gas during  observations. We note that the fraction of 2.0\% is a lower limit and  depends on observation sensitivity. Larger fraction of CO-dark gas will be determined when the spectral sensitivity of CO-dark tracer (e.g., OH and \ci\ ) \cite{2018ApJS..235....1L} increases and the spectral sensitivity of CO emission decreases. In order to provide a quantitative estimation of CO-dark fraction in extragalactic  observations, we took the $\rho$ Oph molecular cloud as a study sample. The fraction of CO-dark gas, $f\rm_{dark}$, is estimated by the following equation \cite{2013ARA&A..51..207B},

\begin{equation}
\begin{split}
    f_{dark} & = \frac{N(\text{CO-dark } H\rm_2)}{N_{\text{tot}}(H\rm_2)} \\
      & = \frac{N(\text{CO-dark } H\rm_2)}{N(\text{CO-dark } H\rm_2) + N(\text{CO-traced } H\rm_2)},
\end{split}
\label{eq:co-dark}
\end{equation}
in which $N$(CO-dark\ H$_2$)=$N_{tot}$(H$_2$)$-N$(CO-traced\ H$_2$) and $N$(CO-traced\ H$_2$)=$X_{CO}$W(CO(1-0)). The factor of CO intensity compared to \h2\ column density, $X_{CO}$, is adopted as the canonical value of $2 \times 10^{20}\ \mathrm{cm^{-2} (K\ km\ s^{-1})^{-1}}$ in the Milky Way, though it depends on physical properties (e.g., metallicity and UV intensity) \cite{2013ARA&A..51..207B}.  The total column density of \h2, $N_{tot}$(H$_2$), is converted by \ci\ column density $N$(\ci) through $N_{tot}$(H$_2$)=$N$(\ci)/$x$(\ci), where median $x$(\ci) value of $1.8\times 10^{-5}$ in $\rho$ Oph molecular cloud. Finally, the derived median value of $f_{dark}$ is 0.43.  

During the above estimation, the contribution of atomic hydrogen on  $f$ value is neglected. This leads to uncertainty in diffuse regions with $A_V< 1$ mag but affects little for $f$ value since the minimal $A_V$ value during calculations reaches 0.94 mag. As our observational capabilities continue to improve, the use of \ci\ as a tracer will undoubtedly play a pivotal role in unraveling the fraction of CO-dark gas and its impact on galaxy evolution.

\subsection{Evidence of Clumpy Geometry from Non-thermal Line Width Distribution}
\label{subsec:clumpy_discussion}

As shown in Section \ref{subsec:intg-codark}, the spatial distribution of $N$(\ci) generally aligns with that of $N$(\13co) and $N$(\h2), which trace the denser interiors of molecular clouds. This finding is inconsistent with the traditional one-dimensional PDR model, where \ci\ is expected to originate from a thin intermediate-extinction layer ($A_{\rm V}$ $\sim$1\,mag) between C$^+$ and CO. The widespread \ci\ emission observed in the $\rho$ Oph cloud resembles findings in Orion and infrared dark clouds (IRDCs). These observations support a clumpy cloud geometry, where UV photons penetrate low-density interclump regions, producing significant \ci\ emission \cite{2013ApJ...774L..20S, 2014A&A...571A..53B, 2022ARA&A..60..247W,2015ApJ...811...13B}.

Though the abundances of different species can be explained with enhanced cosmic ray ionization of PDR model (see Section \ref{pdr_model}), it can not exclude the clumpy model, which is the other main explanation of \ci\ distribution. Investigating the dynamical distribution (e.g., non-thermal line width) of different molecules provides a new view to check this. 

The line width of each velocity component is derived by adopting Gaussian fitting. To ensure a consistent comparison between \ci\ and $^{13}$CO, all spectra were first convolved to a common velocity resolution, matching the native resolution of the \ci\ data. Due to optically thick and significant self-absorption feature toward $^{12}$CO(1-0), we only adopted Gaussian fitting of the \ci\ , $^{13}$CO, and C$^{18}$O spectra with one or two velocity components. The spectral line profile for each pixel was visually examined for characteristic of self-absorption. Pixels exhibiting these features were excluded from further analysis to ensure the reliability of the derived parameters. To evaluate their differences in turbulence, we have calculated the non-thermal velocity linewidth ($\Delta V _{\rm NT}$), which is defined as \cite{2003ApJ...587..262L}: 
\begin{equation}
    \Delta V_{\rm NT} =\sqrt{\Delta V^2 - \dfrac{8kln2 }{\mu m\rm_{H}}T\rm_k},
\end{equation}
where $\Delta V$ is the observed full width at half maximum (FWHM) of Gaussian function, m$_H$ is the hydrogen mass, $k$ is the Boltzmann's constant. and $\mu$ is the molecular  weight , which is 29 for $^{13}$CO and 12 for \ci.  $T\rm_k$ is the kinetic temperature. We adopt the excitation temperature of $^{12}$CO(1-0) as $T\rm_k$ for both $^{13}$CO and \ci.

\begin{figure*}[ht!]
 \includegraphics[width=0.98\textwidth]{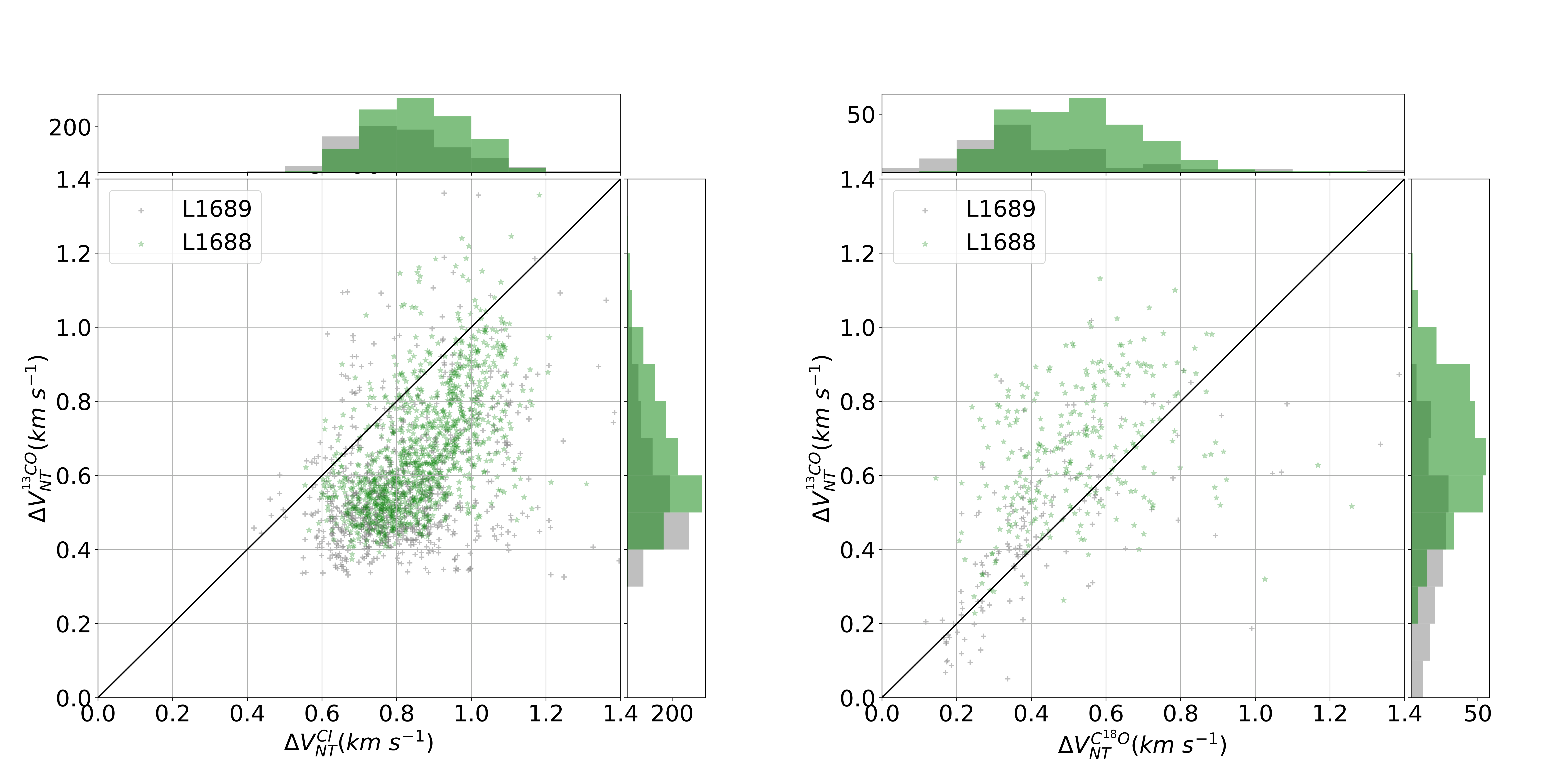}
 \caption{$Left\ Panel$: Distribution of linewidth for $^{13}$CO and \ci. The \ci\ line widths are systematically broader than those of $^{13}$CO, with no apparent linear correlation between the two tracers. $Right\ Panel$: Distribution of linewidth for $^{13}$CO and C$^{18}$O.}
 \label{fig:sigma}
 \end{figure*}

Fig. \ref{fig:sigma} compares the nonthermal line widths ($\Delta V_{\rm NT}$) of \ci, $^{13}$CO and C$^{18}$O. The broader line widths of $^{13}$CO relative to C$^{18}$O suggest that C$^{18}$O traces denser, more shielded gas where turbulence is attenuated \cite{2024RAA....24i5020T}. In contrast, the $\Delta V_{\rm NT}$ values for \ci\ are systematically broader (concentrated between 0.7–1.0 km s$^{-1}$) than those for $^{13}$CO (0.3–0.6 km s$^{-1}$) in both L1688 and L1689, even when accounting for differences in spectral resolution. The poor correlation between the two in a one-to-one comparison indicates that they trace distinct kinematic environments.  We therefore propose that \ci\ resides in a more turbulent, inter-clump medium, while $^{13}$CO is confined to denser, more quiescent clumps \cite{2008A&A...488..623C}. 

Furthermore, while $^{13}$CO exhibits broader lines in the more actively star-forming region L1688 (which hosts six times more YSOs than L1689 \cite{2009ApJS..181..321E}), the \ci\ line widths show no significant regional variation. This discrepancy points to a scenario where stellar feedback selectively injects turbulence into the dense gas traced by $^{13}$CO, leaving the more diffuse, atomic gas largely unaffected.

\section{Summary}\label{sec:summary}

In this study, we conducted a comprehensive analysis of the \r\ Ophiuchus molecular cloud complex, focusing on the large-scale mapping of atomic carbon and its transition to carbon monoxide. Our investigation leveraged data from the Submillimeter Wave Astronomy Satellite (SWAS) and other observatories, providing a detailed view of the spatial distribution, column densities, and abundances of \ci\ and CO in relation to external far-ultraviolet (FUV) intensities and hydrogen molecular column density.

Our data as well as analysis reveal the following points:

\begin{enumerate}
    \item   The spatial distribution of \ci\ traces both extended diffuse regions and dense structures. With current sensitivity of \ci\ emission, we find 2.0\% of the total number of pixels show \ci\ emission with weaker CO emission around the boundary of the $\rho$ Oph molecular cloud (Fig. \ref{fig:CO_dark}). When \ci\ emission is introduced as \h2\ tracer, the CO-dark gas fraction is estimated as 0.43 in this region (section \ref{subsec:co-dark_discussion}). 
    
    \item The transition of \ci\,-to-CO take place at $21.4 < \rm {log_{10}} N_{\rm H_2} < 21.6$, where the abundance of \ci\ gradually decreases as $N_{\rm H_2}$ increases while the abundance of CO shows the opposite trend (Fig. \ref{fig:co_abundance_h2} and Fig. \ref{fig:xci_G0_NH2}).

    \item The abundance of \ci\ decreases gradually with increasing UV intensity but keeps almost constant at high UV intensity ($\rm {log_{10}} G_0 > 2$), and thus \ci\ can serve as a good tracer of H$_2$ in such environments (Fig. \ref{fig:xci_G0_NH2}).  
    
    \item The abundances of different species in $\rho$ Oph cloud can not be explained by PDR model with typical CR ionization rate in the Milky Way (Fig. \ref{fig:CRresults} and Fig. \ref{fig:mcmc_results}).  
    
    \item  Both the spatial distribution and non-thermal line widths between \ci\ and CO support clumpy model (Fig. \ref{fig:sigma}).
    
\end{enumerate}

\textbf{Future Perspective } Current observations of both \ci\ and CO have significantly advanced our understanding of carbon evolution in Galactic molecular clouds. However, two fundamental questions remain, which demand high-sensitivity \ci\ observations across diverse galactic environments:

\begin{enumerate}
\item How do physical properties—such as cosmic-ray flux and metallicity—govern the transition from atomic carbon (\ci\ ) to carbon monoxide (CO) within interstellar clouds?

\item What is the role of stellar feedback (e.g., radiation, stellar winds, and supernovae) in regulating the carbon cycle in molecular clouds?
\end{enumerate}

The upcoming Chinese Survey Space Telescope (CSST), scheduled for launch in 2027, will usher in a new era for astronomy with its state-of-the-art High Sensitivity Terahertz Detection Module. Designed specifically to investigate molecular cloud evolution \cite{2026SCPMA..6939501C}, CSST features a terahertz spectrometer covering 0.41–0.51 THz. It is capable of detecting the \ci\ 492 GHz line with a sensitivity better than 150 mK per 100 kHz at an integration time of 200 seconds \cite{2026SCPMA..6939501C}. This represents a significant improvement over previous missions such as SWAS, positioning CSST as a cornerstone facility for unraveling the lifecycle of carbon in galaxies in the coming decades.

%%%%%%%%%%%%%%%%%%%%%%%%%%%%%%%%%%%%%%%%%%%%%%%%%%%%%%%
%%% Acknowledgements. ??§Ý
%%%%%%%%%%%%%%%%%%%%%%%%%%%%%%%%%%%%%%%%%%%%%%%%%%%%%%%
\Acknowledgements{This work is sponsored by the National Natural Science Foundation of China (grant Nos. 12588202, 12473023, 12373026), the China Manned Space Program with grant no. CMS-CSST-2025-A10, the University Annual Scientific Research Plan of Anhui Province (Nos. 2023AH030052). 
	This work is supported by the Leading Innovation and Entrepreneurship Team of Zhejiang Province of China (Grant No. 2023R01008).
	The research was carried out in part at the Jet Propulsion Laboratory, California Institute of Technology, under a contract with the National Aeronautics and Space Administration (80NM0018D0004).
	Di Li is a New Cornerstone Investigator and is supported by National Key R\&D Program of China No. 2023YFE0110500, grant NSF PHY-2309135 to the Kavli Institute for Theoretical Physics (KITP).}

\InterestConflict{The authors declare that they have no conflict of interest.}

\end{multicols}


\begin{thebibliography}{99}
	\bibitem{1978ApJ...222..881G} Goldsmith, P.~F. \& Langer, W.~D. 1978, \apj, 222, 881
	\bibitem{1991ApJ...377..192H} Hollenbach, D.~J., Takahashi, T., \& Tielens, A.~G.~G.~M. 1991, \apj, 377, 192
	\bibitem{1985ApJ...291..722T} Tielens, A.~G.~G.~M. \& Hollenbach, D. 1985, \apj, 291, 722
	\bibitem{1986ApJS...62..109V} van Dishoeck, E.~F. \& Black, J.~H. 1986, \apjs, 62, 109
	\bibitem{1988ApJ...334..771V} van Dishoeck, E.~F. \& Black, J.~H. 1988, \apj, 334, 771
	\bibitem{2022ARA&A..60..247W} Wolfire, M.~G., Vallini, L., \& Chevance, M. 2022, \araa, 60, 247
	\bibitem{1992IAUS..150..143V} van Dishoeck, E.~F. 1992, Astrochemistry of Cosmic Phenomena, 150, 143
	\bibitem{2005Sci...307.1292G} Grenier, I.~A., Casandjian, J.-M., \& Terrier, R. 2005, Science, 307, 5713, 1292
	\bibitem{2011A&A...536A..19P} Planck Collaboration, Ade, P.~A.~R., Aghanim, N., et al. 2011, \aap, 536, A19
	\bibitem{2013ARA&A..51..207B} Bolatto, A.~D., Wolfire, M., \& Leroy, A.~K. 2013, \araa, 51, 1, 207
	\bibitem{2014A&A...561A.122L} Langer, W.~D., Velusamy, T., Pineda, J.~L., et al. 2014, \aap, 561, A122
	\bibitem{2015ApJ...798....6F} Fukui, Y., Torii, K., Onishi, T., et al. 2015, \apj, 798, 6
	\bibitem{2016A&A...593A..42T} Tang, N., Li, D., Heiles, C., et al. 2016, \aap, 593, A42
	\bibitem{2018ApJS..235....1L} Li, D., Tang, N., Nguyen, H., et al. 2018, \apjs, 235, 1L
	\bibitem{2018A&A...611A..51R} Remy, Q., Grenier, I.~A., Marshall, D.~J., et al. 2018, \aap, 611, A51
	\bibitem{2023A&A...675A.145L} Liszt, H. \& Gerin, M. 2023, \aap, 675, A145
	\bibitem{2020ApJ...889L...4L} Luo, G., Li, D., Tang, N., et al. 2020, \apjl, 889, L4
	\bibitem{1985ApJ...291..747T} Tielens, A.~G.~G.~M. \& Hollenbach, D. 1985, \apj, 291, 747
	\bibitem{1989ARA&A..27...41G} Genzel, R. \& Stutzki, J. 1989, \araa, 27, 41
	\bibitem{1993Sci...262...86T} Tielens, A.~G.~G.~M., Meixner, M.~M., van der Werf, P.~P., et al. 1993, Science, 262, 5130, 86
	\bibitem{1994ApJ...422..136T} Tauber, J.~A., Tielens, A.~G.~G.~M., Meixner, M., et al. 1994, \apj, 422, 136
	\bibitem{1995A&A...294..792H} Hogerheijde, M.~R., Jansen, D.~J., \& van Dishoeck, E.~F. 1995, \aap, 294, 792
	\bibitem{1999RvMP...71..173H} Hollenbach, D.~J. \& Tielens, A.~G.~G.~M. 1999, Reviews of Modern Physics, 71, 1, 173
	\bibitem{2016Natur.537..207G} Goicoechea, J.~R., Pety, J., Cuadrado, S., et al. 2016, \nat, 537, 7619, 207
	\bibitem{2002ApJS..139..467I} Ikeda, M., Oka, T., Tatematsu, K., et al. 2002, \apjs, 139, 2, 467
	\bibitem{2013ApJ...774L..20S} Shimajiri, Y., Sakai, T., Tsukagoshi, T., et al. 2013, \apjl, 774, 2, L20
	\bibitem{2001ApJ...558..176O} Oka, T., Yamamoto, S., Iwata, M., et al. 2001, \apj, 558, 1, 176
	\bibitem{1999sf99.proc...88A} Arikawa, Y., Tatematsu, K., Sekimoto, Y., et al. 1999, Star Formation 1999, 88
	\bibitem{2003ApJ...589..378K} Kamegai, K., Ikeda, M., Maezawa, H., et al. 2003, \apj, 589, 378
	\bibitem{2021ApJ...914L...9Y} Yamagishi, M., Shimajiri, Y., Tokuda, K., et al. 2021, \apjl, 914, L9
	\bibitem{2021PASJ...73..174I} Izumi, N., Fukui, Y., Tachihara, K., et al. 2021, \pasj, 73, 174
	\bibitem{2015ApJ...803...37B} Bisbas, T.~G., Papadopoulos, P.~P., \& Viti, S. 2015, \apj, 803, 37
	\bibitem{2017ApJ...839...90B} Bisbas, T.~G., van Dishoeck, E.~F., Papadopoulos, P.~P., et al. 2017, \apj, 839, 90
	\bibitem{2015MNRAS.450.4424B} Bialy, S. \& Sternberg, A. 2015, \mnras, 450, 4424
	\bibitem{1988ApJ...332..379S} Stutzki, J., Stacey, G.~J., Genzel, R., et al. 1988, \apj, 332, 379
	\bibitem{1990ApJ...365..620B} Burton, M.~G., Hollenbach, D.~J., \& Tielens, A.~G.~G.~M. 1990, \apj, 365, 620
	\bibitem{1993ApJ...405..216M} Meixner, M. \& Tielens, A.~G.~G.~M. 1993, \apj, 405, 216
	\bibitem{1997A&A...323..953S} Spaans, M. \& van Dishoeck, E.~F. 1997, \aap, 323, 953
	\bibitem{2004A&A...424..887K} Kramer, C., Jakob, H., Mookerjea, B., et al. 2004, \aap, 424, 887
	\bibitem{2008A&A...477..547K} Kramer, C., Cubick, M., R{\"o}llig, M., et al. 2008, \aap, 477, 547
	\bibitem{2008A&A...482..197P} Pineda, J.~L., Mizuno, N., Stutzki, J., et al. 2008, \aap, 482, 197
	\bibitem{2008A&A...488..623C} Cubick, M., Stutzki, J., Ossenkopf, V., et al. 2008, \aap, 488, 623
	\bibitem{2021MNRAS.502.2701B} Bisbas, T.~G., Tan, J.~C., \& Tanaka, K.~E.~I. 2021, \mnras, 502, 2701
	\bibitem{2008hsf2.book..351W} Wilking, B.~A., Gagn{\'e}, M., \& Allen, L.~E. 2008, Handbook of Star Forming Regions, Volume II, 351
	\bibitem{2018ApJ...869L..33O} Ortiz-Le{\'o}n, G.~N., Loinard, L., Dzib, S.~A., et al. 2018, \apjl, 869, L33
	\bibitem{2010A&A...518L.102A} Andr{\'e}, P., Men'shchikov, A., Bontemps, S., et al. 2010, \aap, 518, L102
	\bibitem{2015A&A...584A..91K} K{\"o}nyves, V., Andr{\'e}, P., Men'shchikov, A., et al. 2015, \aap, 584, A91
	\bibitem{2020A&A...638A..74L} Ladjelate, B., Andr{\'e}, P., K{\"o}nyves, V., et al. 2020, \aap, 638, A74
	\bibitem{2000ApJ...539L..77M} Melnick, G.~J., Stauffer, J.~R., Ashby, M.~L.~N., et al. 2000, \apjl, 539, L77
	\bibitem{2006AJ....131.2921R} Ridge, N.~A., Di Francesco, J., Kirk, H., et al. 2006, \aj, 131, 2921
	\bibitem{2014A&A...562A.138R} Roy, A., Andr{\'e}, P., Palmeirim, P., et al. 2014, \aap, 562, A138
	\bibitem{2008ApJ...680..428G} Goldsmith, P.~F., Heyer, M., Narayanan, G., et al. 2008, \apj, 680, 428
	\bibitem{2012ApJS..203...13G} Goldsmith, P.~F., Langer, W.~D., Pineda, J.~L., et al. 2012, \apjs, 203, 13
	\bibitem{1981ApJ...251..533P} Phillips, T.~G. \& Huggins, P.~J. 1981, \apj, 251, 533
	\bibitem{1994ARA&A..32..191W} Wilson, T.~L. \& Rood, R. 1994, \araa, 32, 191
	\bibitem{2022RAA....22h5017X} Xia, J., Tang, N., Zhi, Q., et al. 2022, RAA, 22, 085017
	\bibitem{1997MNRAS.287..799I} Ivezic, Z. \& Elitzur, M. 1997, \mnras, 287, 4, 799
	\bibitem{2010ApJ...721..686P} Pineda, J.~L., Goldsmith, P.~F., Chapman, N., et al. 2010, \apj, 721, 686
	\bibitem{2023ApJ...942..101L} Luo, G., Zhang, Z.-Y., Bisbas, T.~G., et al. 2023, \apj, 942, 101
	\bibitem{1998ApJ...499..234C} Caselli, P., Walmsley, C.~M., Terzieva, R., et al. 1998, \apj, 499, 234
	\bibitem{Bisbas12} Bisbas, T.~G., Bell, T.~A., Viti, S., et al. 2012, \mnras, 427, 2100
	\bibitem{McElroy13} McElroy, D., Walsh, C., Markwick, A.~J., et al. 2013, \aap, 550, A36
	\bibitem{Cardelli96} Cardelli, J.~A., Meyer, D.~M., Jura, M., et al. 1996, \apj, 467, 334
	\bibitem{Draine11} Draine, B.~T. 2011, Physics of the Interstellar and Intergalactic Medium, Princeton University Press
	\bibitem{Cartledge04} Cartledge, S.~I.~B., Lauroesch, J.~T., Meyer, D.~M., et al. 2004, \apj, 613, 1037
	\bibitem{Asplund09} Asplund, M., Grevesse, N., Sauval, A.~J., et al. 2009, \araa, 47, 481
	\bibitem{Cazaux02} Cazaux, S. \& Tielens, A.~G.~G.~M. 2002, \apjl, 575, L29
	\bibitem{Cazaux04} Cazaux, S. \& Tielens, A.~G.~G.~M. 2004, \apj, 604, 222
	\bibitem{Cazaux10} Cazaux, S. \& Tielens, A.~G.~G.~M. 2010, \apj, 715, 698
	\bibitem{Bayet09} Bayet, E., Viti, S., Williams, D.~A., et al. 2009, \apj, 696, 1466
	\bibitem{Holdship22} Holdship, J. \& Viti, S. 2022, \aap, 658, A103
	\bibitem{Bisbas23} Bisbas, T.~G., van Dishoeck, E.~F., Hu, C.-Y., et al. 2023, \mnras, 519, 729
	\bibitem{2025A&A...700L..16G} Gaches, B.~A.~L. 2025, \aap, 700, L16
	\bibitem{Dalgarno06} Dalgarno, A. 2006, PNAS, 103, 12269
	\bibitem{Gaches22} Gaches, B.~A.~L., Bisbas, T.~G., \& Bialy, S. 2022, \aap, 658, A151
	\bibitem{Obolentseva24} Obolentseva, M., Ivlev, A.~V., Silsbee, K., et al. 2024, \apj, 973, 142
	\bibitem{Gaches19} Gaches, B.~A.~L., Offner, S.~S.~R., \& Bisbas, T.~G. 2019, \apj, 878, 105
	\bibitem{2010ApJ...716.1191W} Wolfire, M.~G., Hollenbach, D., \& McKee, C.~F. 2010, \apj, 716, 1191
	\bibitem{2015MNRAS.448.1607G} Glover, S.~C.~O., Clark, P.~C., Micic, M., et al. 2015, \mnras, 448, 1607
	\bibitem{1999ApJ...513..275B} Bolatto, A.~D., Jackson, J.~M., \& Ingalls, J.~G. 1999, \apj, 513, 275
	\bibitem{2020A&A...643A.141M} Madden, S.~C., Cormier, D., Hony, S., et al. 2020, \aap, 643, A141
	\bibitem{2014A&A...571A..53B} Beuther, H., Ragan, S.~E., Ossenkopf, V., et al. 2014, \aap, 571, A53
	\bibitem{2015ApJ...811...13B} Burton, M.~G., Ashley, M.~C.~B., Braiding, C., et al. 2015, \apj, 811, 13
	\bibitem{2003ApJ...587..262L} Li, D., Goldsmith, P.~F., \& Menten, K. 2003, \apj, 587, 262
	\bibitem{2024RAA....24i5020T} Tursun, K., Esimbek, J., Baan, W., et al. 2024, RAA, 24, 095020
	\bibitem{2009ApJS..181..321E} Evans, N.~J., Dunham, M.~M., J{\o}rgensen, J.~K., et al. 2009, \apjs, 181, 321
	\bibitem{2026SCPMA..6939501C} CSST Collaboration, Gong, Y., Miao, H., et al. 2026, SCPMA, 69, 239501
\end{thebibliography}
\end{document}